\documentclass[fleqn,usenatbib]{mnras}

\usepackage{newtxtext,newtxmath}

\usepackage[T1]{fontenc}

\usepackage{aecompl}
\usepackage{graphicx}	
\usepackage{amsmath}	
\usepackage{amssymb}	

\usepackage{xcolor}

\usepackage{subfig}
\usepackage{comment}
\usepackage{multirow}

\usepackage{mathrsfs} 

\usepackage{etoolbox}
\makeatletter
\newcount\c@additionalboxlevel
\newcount\c@maxboxlevel
\patchcmd\@combinedblfloats{\box\@outputbox}{%
  \stepcounter{additionalboxlevel}%
  \box\@outputbox
}{}{\errmessage{\noexpand\@combinedblfloats could not be patched}}

\AtBeginShipout{%
  \ifnum\value{additionalboxlevel}>\value{maxboxlevel}%
    \typeout{Warning: maxboxlevel might be too small, increase to %
      \the\value{additionalboxlevel}%
    }%
  \fi 
  \@whilenum\value{additionalboxlevel}<\value{maxboxlevel}\do{%
    \typeout{* Additional boxing of page `\thepage'}%
    \setbox\AtBeginShipoutBox=\hbox{\copy\AtBeginShipoutBox}%
    \stepcounter{additionalboxlevel}%
  }%
  \setcounter{additionalboxlevel}{0}%
}
\makeatother

\newcommand{\quotes}[1]{``#1''}
\newcommand\code[1]{\textsc{\MakeLowercase{#1}}}

\def\be{\begin{equation}}
\def\ee{\end{equation}}

\def\cc{{\rm cm}^{-3}}
\def\kms{km s$^{-1}$}
\def\gnot{{\rm G}_{0}}

\def\msolar{{\rm M}_{\odot}}
\def\msun{{\rm M}_{\odot}}
\def\zsun{{\rm Z}_{\odot}}

\def\msunpc2{\msun/{\rm pc}^{2}}
\def\dsun{\dust_{\odot}}

\def\gsim{\lower.5ex\hbox{\gtsima}} 
\def\lsim{\lower.5ex\hbox{\ltsima}} 
\def\gtsima{$\; \buildrel > \over \sim \;$} 
\def\ltsima{$\; \buildrel < \over \sim \;$} \def\gsim{\lower.5ex\hbox{\gtsima}} 
\def\lsim{\lower.5ex\hbox{\ltsima}} 
\def\simgt{\lower.5ex\hbox{\gtsima}} 
\def\simlt{\lower.5ex\hbox{\ltsima}} 

\def\CII{\hbox{[C~$\scriptstyle\rm II $]}}
\def\OIII{\hbox{[O~$\scriptstyle\rm III $]}}

\def\HII{\hbox{H~$\scriptstyle\rm II\ $}} 

\def\kms{{\rm km\, s^{-1}}}

\definecolor{mkcolor}{HTML}{01abdf} 
\definecolor{apcolor}{HTML}{b3003b}
\definecolor{afcolor}{HTML}{01bdff}

\def\dust{\mathcal{D}}

\def\kms{{\rm km\, s^{-1}}}
\def\nh2{n_{\rm H2}}

\def\sigmant{\sigma_{\rm nt}}
\def\sigmath{\sigma_{\rm th}} 
\graphicspath{{../plots/},{plots/}} 

\title[Velocity dispersion in the ISM of early galaxies]{Velocity dispersion in the interstellar medium of early galaxies}

\author[M. Kohandel et al.]{M. Kohandel$^{1}$\thanks{\href{mailto:mahsa.kohandel@sns.it}{mahsa.kohandel@sns.it}}, A. Pallottini$^{1,2}$, A. Ferrara$^{1}$, S. Carniani$^{1}$, S. Gallerani$^{1}$, L. Vallini$^{3}$,
\newauthor A. Zanella$^{4}$, C. Behrens$^{5}$.
\\
$^{1}$ Scuola Normale Superiore, Piazza dei Cavalieri 7, I-56126 Pisa, Italy\\
$^{2}$ Centro Fermi, Museo Storico della Fisica e Centro Studi e Ricerche \quotes{Enrico Fermi}, Piazza del Viminale 1, Roma, 00184, Italy\\
$^{3}$ Leiden Observatory, Leiden University, PO Box 9500, 2300 RA Leiden, The Netherlands.\\
$^{4}$ INAF - Osservatorio Astronomico di Padova, Vicolo Osservatorio 5, 35122, Padova, Italy\\
$^{5}$ Institut f\"{u}r Astrophysik, Georg-August Universit\"{a}t G\"{o}ttingen, Friedrich-Hundt-Platz 1, 37077, G\"{o}ttingen, Germany\\
}

\date{Accepted XXX. Received YYY; in original form ZZZ}

\pubyear{2020}

\begin{document}
\label{firstpage}
\pagerange{\pageref{firstpage}--\pageref{lastpage}}
\maketitle
%
\begin{abstract}
We study the structure of spatially resolved, line-of-sight velocity dispersion for galaxies in the Epoch of Reionization (EoR) traced by \CII$158\mu\rm{m}$~line emission. Our laboratory is a simulated prototypical Lyman-break galaxy, \quotes{Freesia}, part of the \code{SERRA} suite.
The analysis encompasses the redshift range $6 < z < 8$, when Freesia is in a very active assembling phase. We build velocity dispersion maps for three dynamically distinct evolutionary stages (\textit{Spiral Disk} at $z=7.4$, \textit{Merger} at $z=8.0$, and \textit{Disturbed Disk} at $z=6.5$) using \CII~hyperspectral data cubes.
We find that, at a high spatial resolution of $0.005^{\prime \prime}$ ($\simeq 30 pc$), the luminosity-weighted average velocity dispersion is $\sigma_{\rm{CII}}\simeq 23-38\,\kms$ with the highest value belonging to the  highly-structured Disturbed Disk stage.
Low resolution observations tend to overestimate $\sigma_{\rm CII}$ values due to beam smearing effects that depend on the specific galaxy structure. For an angular resolution of $0.02^{\prime \prime}$ ($0.1^{\prime \prime}$), the average velocity dispersion is $16-34\%$ ($52-115\%$) larger than the actual one.
The \CII~emitting gas in Freesia has a Toomre parameter $\mathcal{Q}\simeq 0.2$ and rotational-to-dispersion ratio of $v_{\rm c}/\sigma \simeq 7$ similar to that observed in $z=2-3$ galaxies.
The primary energy source for the velocity dispersion is due to gravitational processes, such as merging/accretion events; energy input from stellar feedback is generally subdominant ($< 10\%$).
Finally, we find that the resolved $\sigma_{\rm{CII}} - {\Sigma}_{\rm SFR}$ relation is relatively flat for $0.02<{\Sigma}_{\rm SFR}/\msun \rm{yr}^{-1} {\rm kpc}^{-2} < 30$, with the majority of data lying on the derived analytical relation $\sigma \propto \Sigma_{\rm SFR}^{5/7}$. At high SFR, the increased contribution from stellar feedback steepens the relation, and $\sigma_{\rm{CII}}$ rises slightly.
\end{abstract}

\begin{keywords}
galaxies: high-redshift, formation, evolution, ISM -- infrared: general -- methods: numerical
\end{keywords}



\section{Introduction}
The dynamical structure and star formation activity of a galaxy is governed by several interconnected physical processes such as gravity, cooling, heating, feedback, accretion and merging events. As the relative importance of these processes depends on time and environment, the resulting galactic structure/dynamics might be widely different and can be used to study the underlying shaping forces. 

The kpc-scale gas dynamics of star-forming galaxies at $z > 0.5 $ has been massively studied in the literature thanks to groundbreaking observations with Integral Field Unit spectroscopy (IFU, see the review by \citealt{Glazebrook+13}). One of the quantities of interest in such studies is the resolved velocity dispersion, i.e. the line width of emission lines from spatially-resolved observations of the interstellar medium (ISM). IFU observations of galaxies at $1 \le z \le 3$ \citep{Genzel+06, Forster-schreiber+2009, Law+09, Stott+16, Forster-schreiber+18,Mieda+16,Mason+17} have revealed that although a remarkable number of galaxies around the cosmic noon resemble ordered, disk-like structures, they show significantly higher velocity dispersions ($\sim 50-100\, \kms$) compared to local star-forming galaxies ($\sim 20-25\, \kms$, \citealt{Anderson+06,Epinat+10}). The majority of these observations exploit H$\alpha$ and \OIII~lines, i.e. ionized gas tracers. 

The driving mechanism of the observed gas velocity dispersion is a very important and debated issue as it might carry key information on energy deposition and dissipation processes in galaxies. Feedback from star formation activity, including supernovae and radiation pressure, is one of the extensively studied mechanisms (e.g. \citealt{Thompson+05,Lehnert+13, Dib+06, Ostriker+11,LeTrian+11, Shetty+12,Green+14, Martizzi+15,Moiseev+15,pallottini:2017dahlia,Hayward+17,lupi:2019}).

\citet{Green+14}, using H$\alpha$ observations of nearby ($z \sim 0.1$) and intermediate ($1<z<3$) redshift galaxies, have shown that the gas velocity dispersion in a galaxy is correlated with its total star formation rate suggesting that star formation itself is the main driver at all epochs. 
However, except from some analytical works like \citet{Hayward+17}, most theoretical works have struggled to produce velocity dispersions $\simgt 10\,\kms$ purely as a result of stellar feedback (e.g. \citealt{Dib+06, Joung+09, Shetty+12}). Although some models invoke very high momentum input rates to boost the resulting velocity dispersion \citep{Hopkins+11}, it is not yet clear whether such high momentum inputs are physically plausible \citep{Krumholz+12, Rosdahl+15}. 

Alternatives to stellar feedback are different kinds of instabilities occurring on sub-kpc scales. \citet{Kim+03, Piontek+04, Yang+12} have shown that both magneto-rotational and thermal instabilities only produce velocity dispersions of a few $\kms$. Velocity dispersion sourced by the gravitational energy of galaxy-scale accretion flows has also been proposed \citep{Genzel+11}; at present, though, it is unclear whether this mechanism provides enough energy to support the observed velocity dispersion \citep[e.g.][]{Elmegreen+10,Hopkins+13, Klessen+10, Krumholz+16}. 
Despite these efforts, whether observed gas dynamics in local and intermediate-redshift ($z\sim 2$) galaxies is driven by gravitational processes \citep[e.g.][]{Orr+19} or stellar feedback \citep{Genzel+11} is still debated.

Thanks to the Atacama Large Millimeter Array (ALMA), our knowledge on dynamics and structure of early galaxies, deep into the Epoch of Reionization (EoR, $z>6$), is rapidly expanding (for a recent review, see \citet{Dayal18}. While galaxies at EoR have been discovered with UV surveys \citep{Smit+18, Bouwens+15}, far-infrared line observations are crucial to gain information on the physical and dynamical properties of these systems. The \CII$158\mu\rm{m}$~line emission of singly ionized carbon  is one of the strongest coolants of the ISM; it is observed in both individual sources \citep[e.g.][]{Capak+15,Maiolino+15, Pentericci+16,Carniani+17,Jones+17, Matthee+17,Carniani+18himiko,Carniani+18clumps,Smit+18,Harikane:2019}, and large galaxy samples, such as the ALPINE survey \citep{Alpine+19}.

The \CII~emitting gas in galaxies often shows very complex patterns. Multi-component and clumpy structures are common features of these high redshift systems \citep{Carniani+18clumps}. Dynamically, the \CII~emission is interpreted as arising either from rotating disks \citep{Smit+18} or mergers \citep{Jones+17}. This diversity is more notable in large surveys like ALPINE \citep{Alpine+19}, where evidence for a significant number of mergers (almost $40\%$), dispersion dominated disks ($20\%$),  and rotating disks ($14\%$) has been reported.

Forthcoming high angular resolution data will allow us to study in detail the dynamical properties of such systems, answering the following fundamental questions:
\begin{itemize}
\item[1.] How large is the velocity dispersion in EoR galaxies compared to intermediate- and low-$z$ systems?
\item[2.] What is the energy source powering the gas velocity dispersion in EoR galaxies?
\item[3.] Is there any correlation between the spatially resolved velocity dispersion and star formation rates?
\end{itemize}

These questions can be addressed by using high-resolution simulations of galaxies at the EoR. In the last few years, theoretical efforts have attempted to model FIR line emissions to interpret the total luminosity of galaxies observed at $z \ge 6$ and estimate the relative contribution from different phases of the ISM \citep[][]{vallini:2013, vallini:2015, Olsen+17,pallottini:2017dahlia, Katz+19, pallottini:2019,ferrara:2019, Arata20}. These works agree on the fact that most of the \CII~luminosity is produced in Photo Dissociation Regions (PDRs, \citep{pallottini:2017dahlia} with a weak dependence on galaxy mass \citep{Olsen+17}. More recently, attention is turning to model and interpret kinematical observables such as \CII~integrated line profiles \citep{Kohandel+19} using high-resolution simulations. In \citet{Kohandel+19}, we showed that EoR galaxies are actively assembling and developing structures similar to those observed so far \citep{Smit+18, Jones+17, Alpine+19}. These structural/morphological differences, corresponding to rotating disks, mergers or disturbed disks
imprint unique and distinguishable features in the \CII~line spectrum. 

In this work, we want to extend these theoretical studies by modeling dynamical observables using state-of-the-art zoom-in simulations of galaxies at the EoR. These observables include the 2D spatially-resolved mean velocity and velocity dispersion maps derived from hyperspectral data cubes. To this aim, we bridge advanced zoom-in galaxy simulations to IFU-like observations (Sec. \ref{sec_bridge}) by modeling the data cubes for \CII~line emission. We use \code{SERRA}, a suite of zoom-in simulations of EoR galaxies presented in Sec. \ref{sec-serra}. For our dynamical studies, we choose three evolutionary stages of one of these galaxies (called \quotes{Freesia}), i.e. Spiral Disk, Merger and Disturbed Disk (Sec. \ref{sec-stages}). Since the main focus of this paper is to understand ISM velocity dispersion, in Sec. \ref{sec-dispersion} we analyze spatially-resolved velocity dispersion maps extracted from \CII~data cubes for different dynamical stages, and then identify the physical drivers of the velocity dispersion. Finally, in Sec. \ref{sec-sigma-sfr}, we investigate the relation among different components of the velocity dispersion and star formation rate\footnote{We assume cosmological parameters compatible with \citet{Planck+14}, i.e. total vacuum, matter, and baryonic densities in units of the critical density $\Omega_{\Lambda}= 0.692$, $\Omega_{m}= 0.308$, $\Omega_{b}= 0.0481$, Hubble constant $\rm H_0=100\,{\rm h}\,{\rm km}\,{\rm s}^{-1}\,{\rm Mpc}^{-1}$ with ${\rm h}=0.678$, spectral index $n=0.967$, $\sigma_{8}=0.826$.}.
 
\section{SERRA: simulating galaxies in the EoR}\label{sec-serra}

Full details of the \code{SERRA} suite of zoom-in simulations of galaxies at the EoR are described in \citet{pallottini:2019}. 
In the \code{SERRA} suite, a customized version of adaptive mesh refinement (AMR) code \code{RAMSES} \citep{Teyssier+02} is used to evolve gas and dark matter.
Concerning the chemistry, \code{KROME} \citep{Grassi+14} is used to generate a chemical network, in order to follow the non-equilibrium chemistry of H$_2$ \citep{bovino:2016aa,pallottini:2017althaea}, that in turn is converted into stars with a Kennicutt-Schmidt (KS, \citealt{schmidt:1959,Kennicutt+98}) relation. The selected chemical network includes H, H$^{+}$, H$^{-}$, He, He$^{+}$, He$^{++}$, H$_{2}$, H$_{2}^{+}$ and electrons, for a total of about 40 reactions \citep{bovino:2016aa}.
Metallicity ($Z$) is tracked as the sum of heavy elements, assuming solar abundance ratios of different metal species \citep{asplund:2009}.
Dust evolution is not explicitly tracked during the simulation. In \code{SERRA}, it is assumed that the dust-to-gas mass ratio scales with metallicity, i.e. $\dust = \dsun (Z/\zsun)$ -- where $\dsun/\zsun \simeq 0.3$ for the Milky Way (MW) \citep[][]{Hirashita+02} -- and a MW-like grain size distribution is adopted \citep{weingartner:2001apj}. An initial metallicity floor $Z_{\rm floor}=10^{-3}\zsun$ is adopted, as expected from a pre-enrichment scenario in the circumgalactic and intergalactic medium of cosmic density peaks \citep[][]{Madau01, pallottini+14,pallottini:2014}.

The interstellar radiation field (ISRF) is tracked on-the-fly using the moment-based radiative transfer code \code{RAMSES-RT} \citep{Rosdahl+13}, that is coupled to the chemical evolution of the gas \citep{pallottini:2019,decataldo:2019}; in \code{SERRA} the speed of light is reduced by a factor of 100; 5 energy bins are tracked: one partially covering the Habing band ($6.0<{h}\nu <11.2$), one to follow Lyman-Werner band ($11.2<{h}\nu <13.6$) to account for H$_2$ photoevaporation, and 3 to cover ionization of H up to up to the first ionisation level of He ($13.6<{h}\nu <24.59$).

In summary, the simulations start at $z=100$ from cosmological initial conditions generated with \code{music} \citep{hahn:2011mnras}. Then at $z\simeq 6$ the simulations zoom on the DM halo which is hosting the targeted galaxy. The total simulation volume is $(20\,{\rm Mpc}/{\rm h})^{3}$, and it is evolved with a base grid with 8 levels (gas mass $6\times 10^6\msun$). The zoom-in region has a volume of $(2.1\,{\rm Mpc}/{\rm h})^{3}$ and is resolved with 3 additional levels of refinement, thus yielding a gas mass resolution of $m_b = 1.2\times 10^4 \msun$. In this zoom-in region, we allow for 6 additional levels of refinement based on a Lagrangian-like criterion. This enables us to reach scales of $l_{\rm res}\simeq 30\,{\rm pc}$ at $z=6$ in the densest regions, i.e. the most refined cells have mass and size typical of Galactic molecular complexes \citep[MC, e.g.][]{Federrath+13}. In this work, we focus our analysis on  \quotes{Freesia}, a prototypical Lyman-break galaxy in the \code{SERRA} suite.

\subsection{Star formation and stellar feedback}\label{sec:feedback}

In \code{SERRA}, the star formation rate density ($\dot{\rho}_{\star}$) depends on the H$_2$ density ($\rho_{\rm H2}$) via a \citet{schmidt:1959,Kennicutt+98}-like relation:
\be
\dot{\rho}_{\star} = \zeta_{\rm sf} \frac{\mu {\rm m}_{\rm p} \nh2}{t_{\rm{ff}}}\,,
\ee
where $\dot{\rho}_{\star}$ is the local star formation rate density, $\zeta_{\rm sf}$ the star formation efficiency, $\rm m_p$ the proton mass, $\mu$ the mean molecular weight, and $t_{\rm ff}$ the free-fall time.
The star formation efficiency is set to $\zeta_{\rm sf}=10\%$, by adopting the average value observed for MCs \citep{murray+11}, while molecular hydrogen density $n_{\rm H2}$ computation is included in the non-equilibrium chemical network.
As shown in \citet{pallottini:2017dahlia}, the adopted SFR prescription gives similar results to other schemes in which the efficiency is derived from a turbulent virial theorem criterion \citep{semenov:2015}.

A single star particle in \code{SERRA} can be considered as a stellar cluster, with metallicity $Z_{\star}$ set equal to that of the parent cell. For the stellar cluster, a \citet{Kroupa+11} initial mass function is assumed. By using \code{starburst99} \citep{Leitherer+99}, single population stellar evolutionary tracks given by the {\tt padova} \citep{padova} library are adopted, that covers the $0.02 \leq Z_{\star}/\zsun \leq 1$ metallicity range.

In \code{SERRA}, we account for stellar energy inputs and chemical yields that depend both on metallicity $Z_\star$ and age $t_\star$ of the stellar cluster. Stellar feedback includes supernovae (SNe), winds from massive stars, and radiation pressure. Due to the stellar feedback, gas elements of the ISM perceive pressure in both thermal ($\rm{P}_\mathrm{th}$) and non-thermal ($\rm{P}_\mathrm{nt}$) forms. The detailed description of thermal and non-thermal pressure terms due to  stellar feedback can be found in \citet{pallottini:2017dahlia}. The non-thermal pressure mimics the stellar feedback-driven turbulence \citep{Agertz+13,Agertz+15,teyssier:2013mnras}. So these pressure terms induce random gas motions that we define as the thermal ($\sigmath$) and turbulent ($\sigmant$) velocity dispersion:
\be \label{eq:int_disp}
\sigmath = \sqrt{\frac{\rm{P}_\mathrm{th}}{\rho}} , \qquad
\sigmant = \sqrt{\frac{\rm{P}_\mathrm{nt}}{\rho}}
\ee
where $\rho$ is the total gas density in the cell.
The thermal component is affected by gas cooling processes, while the turbulent component dissipates with a time scale given by the eddy turn-over time \citep{MacLow+1999}:
\begin{equation}\label{eq:dissipation}
 t_{\rm diss} \simeq 0.9 \left(\frac{l_{\rm cell}}{10\, {\rm pc}}\right) \left(\frac{10\, \kms}{\sigmant}\right) {\rm Myr}\,,
\end{equation}
where $l_{\rm cell}$ is the size of the cell.

{As detailed in \citet{pallottini:2017dahlia}, stellar feedback incorporates Type II and Ia SNe, winds from OB and AGB stars, and radiation pressure. The energy dissipation in MCs for SN blastwaves \citep{ostriker+88} and OB/AGB stellar winds \citep{Weaver+77} is also accounted \citep[][see in particular Sec.~2.4 and App. A]{pallottini:2017dahlia}.
Continuous mechanical energy deposition rate from winds and supernovae is derived from the stellar tracks, and added in the cell where the star resides. The relative fraction of thermal and kinetic energy depends on the SN blast stage: energy conserving Sedov-Taylor stage (about 70\% thermal, 30\% kinetic), shell formation stage, and pressure driven snowplow (about 15\% thermal and 35\% kinetic).

For radiation pressure, the kinetic energy is computed from the momentum injection rate, in turn based on the luminosity of the source and the optical thickness of the gas to the radiation in various bands \citep[e.g.][]{Krumholz+12}. We use an energy-based implementation that mimics that typically adopted in particle-based codes \citep{Hopkins+11}.

Stellar tracks are also used to calculate photon production. As shown in \citet[][in particular see Fig. 1 therein]{pallottini:2019} at each time step, stars dump photons in the hosting cell in each energy bin according to their stellar age and metallicity. Photons are then advected and absorbed in the radiation step, contributing at the same time to the photo-chemistry.
Dust and gas account for absorption of the radiation, consistently with the chemical reaction cross sections and the \citet{weingartner:2001apj} dust distribution \citep[see][in particular Fig. 2 therein]{pallottini:2019}.
Note that, at a given halo mass, \code{SERRA} galaxies feature star formation and stellar mass histories that are consistent with \citet{lupi:2020MNRAS}, which uses a set of feedback prescriptions similar to \code{FIRE2} \citep{Hopkins+18}.
}

For the analysis of the simulation, it is convenient to define the star formation rate surface density ($\Sigma_{\rm SFR}$) as
\be \label{eq:sfr}
\Sigma_{\rm SFR} = \frac{\Sigma_{\star}(t_\star < \Delta t)}{\Delta t}\,,
\ee
where we account for young star clusters i.e. setting $\Delta t < 30\, \rm{Myr}$.

\section{Bridging simulations and IFU observations}\label{sec_bridge}

\subsection{Modelling \CII~line emission}

The chemical network used in \code{SERRA} includes $\rm{H}$, $\rm{He}$, $\rm{H}^+$, $\rm{H}^-$, $\rm{He}$, $\rm{He}^+$, $\rm{He}^{++}$, $\rm{H}_2$, $\rm{H}_2^{+}$ and electrons. The abundance of other metals are calculated by assuming solar abundances. Ion abundances (e.g. $\rm{C}^+$) and corresponding line emission are computed in post-processing on a cell by cell basis. We use the spectral synthesis code, \code{CLOUDY} \citep{Ferland+17} to predict the \CII~line emission and $\rm{C}^+$ ion. 
The resolution of \code{SERRA} suite is $\simeq 30 \,\rm {pc}$. Hence, we do not resolve the internal structure of molecular clouds, that is sub-parsec scales; also as noted in \citet{pallottini:2019}, gas cell is typically optically thick to ionizing radiation, i.e. for a typical ionized region with ionization parameter $U\sim 10^{-2}$, metallicity $Z=0.5\zsun$, and density $n\simeq 300 {\rm cm}^{-3}$, the resulting \HII~regions size is about $1\,\rm{pc}$, where sub-pc resolutions would be needed to resolve it (see also \citealt{decataldo:2019}, in particular Fig. 4 therein). To overcome these limitations, we have adopted the same post-processing model of \citet{pallottini:2019}, summarized below.

Emission coming from the small scale clumps inside the molecular clouds is accounted similarly to \citet{vallini:2017,vallini:2018}; each molecular cloud with a volume $V$ encompasses clumps with sizes of the Jeans length ($l_J$) and it is characterized by a differential number of clumps ${\rm d}N_{\mathrm{clump}}$:
\begin{subequations}\label{eq_clump_distribution}
\begin{equation}
{\rm d}N_{\mathrm{clump}} = (V/l_J^3){\rm d}P\,,
\end{equation}
where ${\rm d}P$ is the distribution of density $n$ inside a molecular cloud with mean density $n_{0}$; ${\rm d}P$ can be described via a log-normal function \citep{padoan+11}
\begin{equation}
{\rm d}P = \frac{1}{\sigma_s\sqrt{2\pi}}\exp{-\left(\frac{s-s_0}{\sigma_s\sqrt{2}}\right)^2}{\rm d}s \, ,
\end{equation}
with $s$ being the normalized density $s=\ln{(n/n_0)}$, $s_0 \equiv - 0.5 \sigma_s^2$, and $\sigma_s$ being the standard deviation of the distribution; the latter depends on the Mach number ($\mathcal{M}$) as \citep{Krumholz+05}
\begin{equation}
\sigma_s^2 = \ln{(1+(\mathcal{M}/2)^2)}\,.
\end{equation}
We can compute the Mach number using the thermal and non-thermal pressure terms for each cell of gas obtained self-consistently from the simulation (eq. \ref{eq:int_disp})
\begin{equation}
\mathcal{M} = \sqrt{1+\frac{P_{\rm nt}}{P_{\rm th}}}\,.
\end{equation}
\end{subequations}

As in \citet{pallottini:2019}, to compute the \CII~emission we build two grids of \code{CLOUDY} models, i.e. with and without ionising radiation. Every grid is divided in seventeen bins of number density ($10^{-2}  \le n/\cc \le 10^{6.5}$), eight bins of metallicity ($10^{-3} \le Z/\zsun \le 10^{0.5}$) and twelve bins of ISRF ($10^{-1} \le G/\gnot \le 10^{4.5}$), for a total of 1632 distinct models per each grid.
For the spectral energy distribution (SED) of the impinging radiation field on the slab of gas of interest in \code{CLOUDY}, we use a SED taken from \code{STARBURST99} \citep{Leitherer+99} with stellar age of $10\,\mathrm{Myr}$ and solar metallicity. Such stellar population is the primary contributor to the interstellar radiation field in our simulated galaxies\footnote{Note that while in the simulation stellar radiation is tied to stellar metallicity and age, achieving the same result via a grid of \code{cloudy} models is unfeasible. This would require one grid dimension per radiation bin, whose adopted number is typically $\approx 1000$.}.
The intensity of the radiation field is rescaled with the local $G$ flux; if the simulated cell has an ionization parameter $U>10^{-4}$ or if it contains young stars ($t_\star \le 10\, \mathrm{Myr}$), we use the grid with ionising radiation. Otherwise we use the intensity obtained computed without the ionising radiation.
For each clump inside a cell of the simulation, given the input parameters ($n$, $G$, $Z$ and $N$), we compute the \CII~line luminosity per unit area ($\mathscr{L}_{\mathrm{\rm clump}}^{\rm{[CII]}}$) and $C^+$ ion mass ($\mathscr{M}_{\mathrm{clump}}^{C^+}$) by interpolating the values evaluated by \code{CLOUDY} grids.
Then, to account for the cloud structure, we integrate the the clump distribution (eq.s \ref{eq_clump_distribution}) to obtain the total \CII~luminosity ($L_i^{\rm{[CII]}}$) and $C^+$ ion mass ($M_i^{C^+}$) of the i-th cell as:
\begin{subequations}
\begin{equation}\label{eq-cii-em}
L_i^{\rm{[CII]}} = \int \mathscr{L}_{\mathrm{clump}}^{\CII} l_J^2 {\rm d}N_{\mathrm{clump}}\,,
\end{equation}
\begin{equation}
M_i^{C^+} = \int \mathscr{M}_{\mathrm{clump}}^{C^+} l_J^2 {\rm d}N_{\mathrm{clump}} \,.
\end{equation}
\end{subequations}

\subsection{Hyperspectral Data cubes}\label{sec:data_cubes}

\begin{figure*}
\centering
\includegraphics[width=0.9\textwidth]{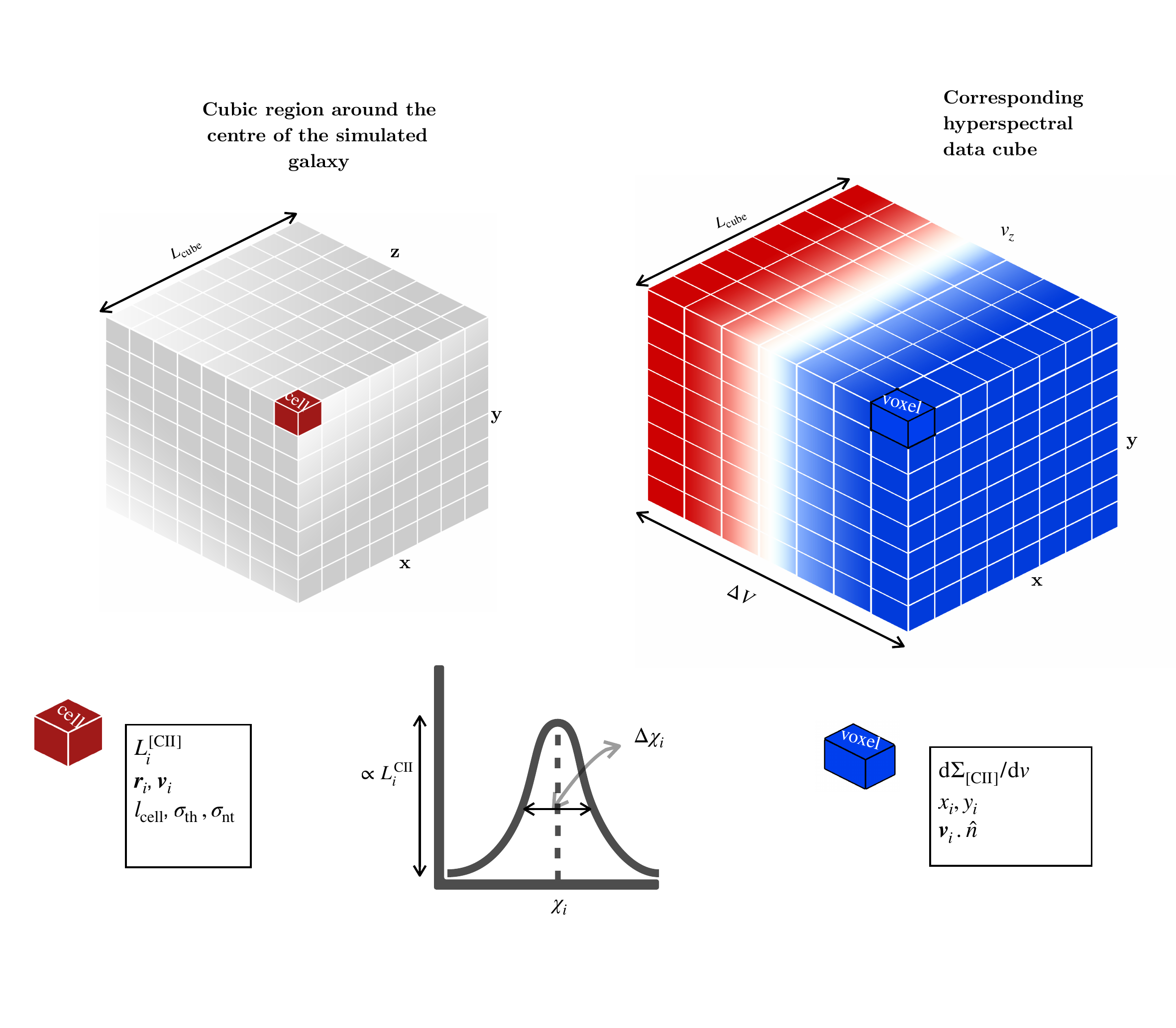}
\caption{Sketch of the model used in this work to obtain hyperspectral data cubes from a simulation box. Having the information on the \CII~luminosity ($L_{i}^{\rm{CII}}$), positions and velocities of the gas particles along with the cell size($l_{\rm{cell}}$), thermal ($\sigma_{\rm{th}}$) and non-thermal($\sigma_{\rm{nt}}$) line broadenings in a simulated cube, one can construct the hyperspectral data cubes with two spatial and one spectral dimensions. The mapping from the simulated box to the hyperspectral cube is done with three Gaussian filters in each dimension. The simulation box is represented as a uniformly binned space only for visualization purposes. See Sec. \ref{sec:data_cubes} for the details. \label{fig:sketch}}
\end{figure*}

Interferometric observations like those obtained with ALMA, VLA, and LOFAR or Integral Field Units like SINFONI and MUSE yield multi-channel data cubes. These data cubes have two spatial dimensions ($x$ and $y$) and one spectral dimension ($\lambda$). These data are sometimes called hyperspectral since they have an extremely high spectral resolution (e.g few thousands of frequency channels in the case of ALMA). The frequency dimension of these data cubes which can be translated into a line-of-sight (l.o.s.) velocity information, allows the observer to study the dynamics of the galaxies even at very high redshifts.

To have a fair comparison between observations and simulations, we generate the so-called Hyperspectral Data Cubes (hereafter HDC) for our simulated galaxies, and extract dynamical observables (see Sec. \ref{sec:model_moments}).
A sketch of the model is given in Fig. \ref{fig:sketch} and the process is detailed below.

First we extract a cubic region around the center of the galaxy with a side-length $L_{\rm{cube}}$ containing a number $N_{\rm cell}$ of AMR cells. For each gas cell we have information on its position ($\textbf{r}_i$), velocity ($\textbf{v}_i$), and \CII~luminosity ($L_i^{\rm{[CII]}}$, computed in post-processing using eq. \ref{eq-cii-em}).
Our HDC has two spatial and one velocity dimensions. Let us call $z$ the l.o.s. direction, so that $v^z = {\bf v} \cdot \hat{\bf z}$ is the velocity component parallel to the l.o.s. and $x-y$ is the plane perpendicular to it.
Then, the velocity-dependent \CII~surface brightness for each voxel of coordinates ($x$, $y$, $v^{z}$) can be modelled as follows: \begin{subequations}\label{eq:data_cube}
\be
\frac{{\rm d}\Sigma_{\rm [CII]}}{{\rm d}v}(x,y,v^{z}) = \sum_{i=1}^{N_{\rm cell}} L_i^{\rm [CII]} K(x,x_i,\Delta x_i) K(y,y_i,\Delta x_i) K(v^z,{v^{z}}_i,\Delta v_i)\,,
\ee
where
\be\label{eq:kernel}
K(\chi,\chi_i, \Delta \chi_i) = \frac{1}{\Delta \chi_i\sqrt{2 \pi}} \exp{-\left(\frac{\chi-\chi_i}{\Delta \chi_i}\right)^2}
\ee
\end{subequations}
represents the general Gaussian kernel\footnote{We have tested different kernels, finding no appreciable differences in the resulting observables.} adopted for three dimensions (2 spatial + 1 velocity). The width of the Gaussian kernels ($\Delta \chi_i$) for spatial dimensions is $\Delta x_i = \Delta y_i =l_{\rm{cell}}$, with $l_{\rm{cell}}$ being the size of the considered cell. For the spectral dimension, $\Delta v_i = (\sigmath^2 + \sigmant^2)^{1/2}$, where $\sigmath$ and $\sigmant$ denote the thermal and non-thermal line broadening, respectively (see Eq. \ref{eq:int_disp}).

\subsection{Line spectrum and emission moment maps}\label{sec:model_moments}

Having the HDC, observables such as the line spectrum and various moments of the specific \CII~surface brightness ${\rm d}\Sigma_{\rm [CII]}/{{\rm d}v}$ can be obtained. It is useful to label $n_x$, $n_y$ and $n_{v}$ the number of bins in each dimension of the HDC, such that ${\Delta}x$, ${\Delta}y$, and ${\Delta}v$ are the corresponding spatial and spectral resolutions.
The integrated $1$D line spectrum can be defined from the HDC as:
\be\label{def_spectrum_integrated}
\frac{{\rm d}L_{\rm [CII]}}{{\rm d}v}(v^z) = \sum_{l=1,m=1}^{n_x,n_y} \frac{{\rm d}\Sigma_{\rm [CII]}}{{\rm d}v}(x_l, y_m, v^z) {\Delta}x\, {\Delta}y\, .
\ee
The integrated surface brightness ($\Sigma_{\rm{[CII]}}$) as well as dynamical observables such as spatially resolved mean velocity ($\langle v \rangle$) and velocity dispersion ($\sigma_{\rm{CII}}$) maps are obtained from the velocity moments of the HDC as follows:
\begin{subequations}\label{eq:moments}
\begin{align}
\Sigma_{\rm{[CII]}}(x,y) &= \sum_{j=1}^{n_{v}} \frac{{\rm d}\Sigma_{\rm [CII]}}{{\rm d}v}(x, y, v^z_j) {\Delta}v\,, \label{def_moment1}\\
\langle v \rangle (x,y)  & = \frac{1}{\Sigma_{\rm{[CII]}}(x,y)}\sum_{j=1}^{n_{v}} v^z_j \frac{{\rm d}\Sigma_{\rm [CII]}}{{\rm d}v}(x, y, v^z_j) {\Delta}v\,, \label{eq:mean_vel}\\
\sigma^2_{\rm{CII}}(x,y) & =  \frac{1}{\Sigma_{\rm{[CII]}}(x,y)} \sum_{j=1}^{n_{v}}  \left(v^z_j-\langle v \rangle (x,y)\right)^2 \frac{{\rm d}\Sigma_{\rm [CII]}}{{\rm d}v}(x, y, v^z_j) {\Delta}v \,. \label{eq:vel_disp_cii}
\end{align}
\end{subequations}

\subsection{Numerical setup for the hyperspectral data cubes}

The HDCs produced in this work have the following setup.
We select a cubic region centered on Freesia with side-length of $L_{\rm{cell}} = 8\,\rm{kpc}$, which typically contains $N_{\rm{cell}} \sim 10^7$ AMR gas cells; for both spatial dimensions we use $n_x = n_y = 256$.
The l.o.s. velocities depend on the inclination of the galaxy; we use $n_{v}=256$ bins to map a $(-400, +400)\, \kms$ velocity range, that is centered on the peak of the \CII~emission.
Thus, the resulting hyperspectral data cubes have $N_{\rm voxel} = 256^3$ voxels, with a spectral resolution of ${\Delta}v \simeq 3.1\,\kms$ and a spatial resolution of ${\Delta}x = {\Delta}y \simeq 31.2 \, \rm{pc}$; the latter corresponds to an angular of $0.005^{\prime\prime}$ at $z=6$.
To speed up the computation of the HDCs, in eq. \ref{eq:kernel} we set the kernel to zero beyond 5 standard deviations away from the mean, i.e. $K(\chi, \chi_i, \Delta \chi_i)=0$ when $|\chi - \chi_i | > 5 \Delta \chi_i$.

\section{Identification of dynamical stages}\label{sec-stages}

\begin{table*}
\caption{Properties of different evolutionary stages of Freesia depicted in Fig. \ref{fig:cii_moment_maps}.
\textit{Notes}: $^\dag$: burstiness parameter defined in eq. \ref{eq_ks}, estimated with the total SFR and gas mass within $r_d$ = 1 kpc; $^\ddagger$: gas fraction defined as $f_g=\Sigma_g/(\Sigma_g+\Sigma_\star)$.
\label{tab:stages_properties}
}
\begin{center}
\begin{tabular}{|l|c|c|c|c|c|c|c|c|c|}
\hline
Stages  &  Short name & redshift & $\rm{M}_{\star}$& $\rm{M}_{\rm{g}}$ & $\rm{SFR}$ & $\rm{L}_{\rm{[CII]}}$  & {$v_{\rm c}$} & $k_s^\dag$ & $f_g^\ddagger$\\ 
~ & ~ & ~ & $[10^{9}\msolar]$ & $[10^{9}\msolar]$ & $[\msolar \rm{yr}^{-1}]$ & $[10^{8}\rm{L}_{\odot}]$      & $[\rm km/s]$  & ~ & ~          \\
\hline
Spiral Disk    & SD &$7.4$     & $4.9$                          & $3.4$ & $38.4$ & $1.0$                     & {$189$}       & 2.7 & 0.42      \\
Merger         & MG &$8.0$     & $4.0$                          & $3.0$ & $29.5$ & $0.7$                     & {$173$}       & 2.6 & 0.43       \\
Disturbed Disk & DD &$6.5$     & $10.5$                         & $3.6$ & $85.4$ & $1.6$                     & {$246$}       & 5.6 & 0.26      \\
\hline
\end{tabular}
\end{center}
\end{table*}
\begin{figure*}
\centering
\includegraphics[width=0.9\textwidth]{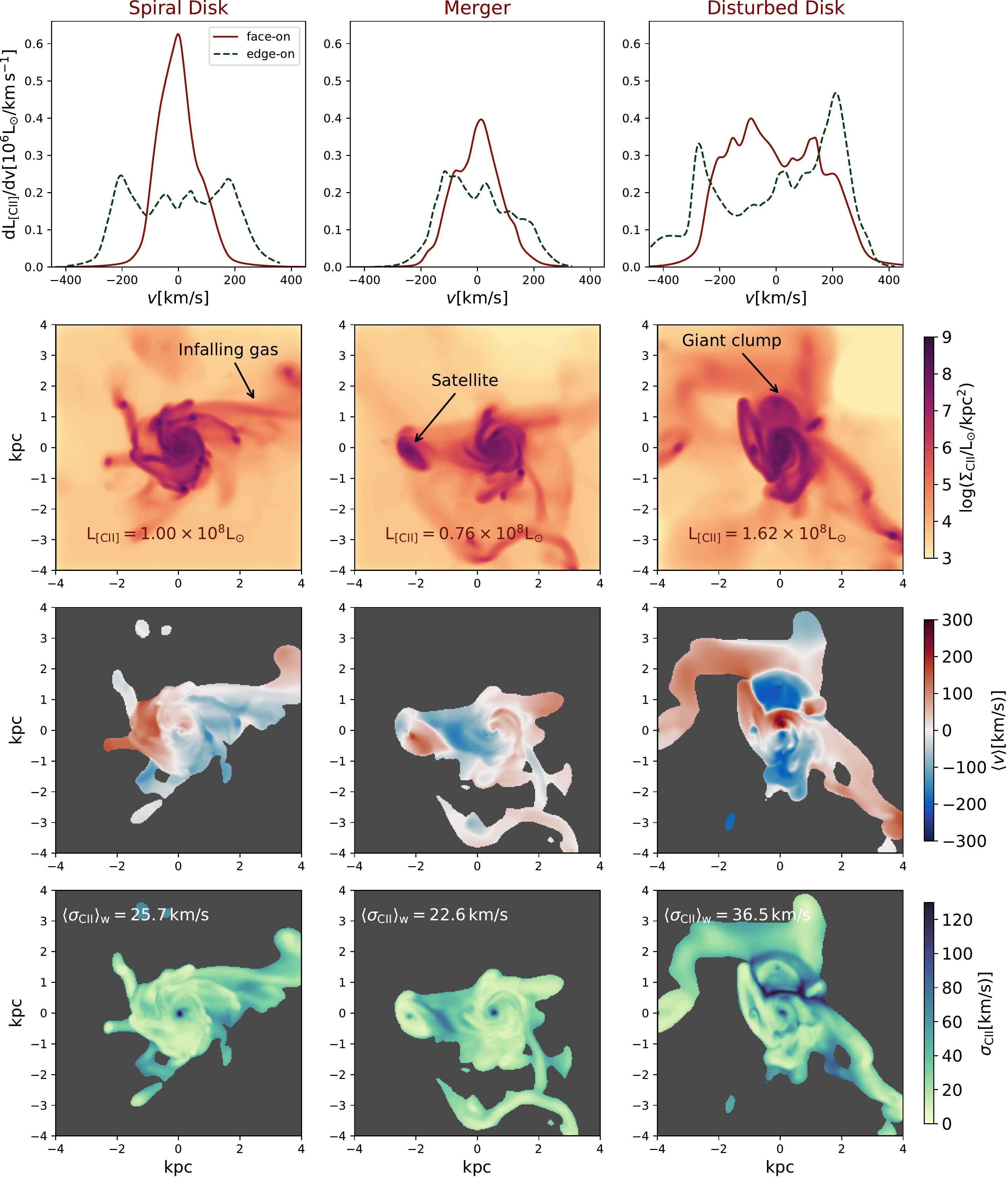}
\caption{Observables derived from \CII~line hyperspectral data cubes for three stages of Freesia, Spiral Disk (SD, {\it left panels}), Merger (MG, {\it middle panels}) and Disturbed Disk (DD, {\it right panels}).
From the top the plotted quantities are: $1$D \CII~line spectra (${\rm d}L_{\rm [CII]}/{\rm d}v$, {\it first row}), \CII~surface brightness ($\Sigma_{\rm{[CII]}}$, moment-0 map, {\it second row}), mean velocity ($\langle v\rangle$, moment-1 map, {\it third row}), and velocity dispersion ($\sigma_{\rm [CII]}$, moment-2 map {\it fourth row}).
Spectra are extracted for face-on and edge-on views, maps are shown for the face-on view.
In the surface brightness maps we report the total luminosity as an inset; on the velocity dispersion maps we quote $\langle \sigma \rangle_{\rm{w}}$, the \CII~ luminosity weighted average of velocity dispersion. In the mean velocity and dispersion maps we gray out pixels with $\Sigma_{\rm CII} < 10^{4.5}L_{\odot}/\rm{kpc}^2$.
\label{fig:cii_moment_maps}}
\end{figure*}

\begin{table*}
\caption{List of definition and symbols used for the various components of the velocity dispersion. \label{tab:sigma_list}}
\begin{tabular}{|l|l|l|l|}
\hline
Symbol               & Description                                                         & Expression                                                 & Reference \\\hline
$\sigma_{\rm{th}}$   & Thermal velocity dispersion due to stellar feedback                 & $\sigmath = \sqrt{ {P}_\mathrm{th}/\rho }$        & eq. \ref{eq:int_disp} \\
$\sigma_{\rm{nt}}$   & Non-thermal (turbulent) velocity dispersion due to stellar feedback & $\sigmant = \sqrt{ {P}_\mathrm{nt}/\rho }$        & eq. \ref{eq:int_disp} \\
$\sigma_{\mu}$       & Total (small-scale) velocity dispersion due to stellar feedback             & $\sigma_{\mu}^2 = \sigma_{\rm{nt}}^2 + \sigma_{\rm{th}}^2$ & eq. \ref{eq:sigma_mu_def} \\
$\sigma_{\rm{[CII]}}$& Velocity dispersion derived from moment-2 of \CII~line emission     & $\sigma_{\rm{[CII]}}^2 = \sigma_\mu^2 + \sigma_{\rm{b}}^2$ & eq. \ref{eq:vel_disp_cii}\\
$\sigma_{\rm{b}}$    & Velocity dispersion due to bulk motions                             & $\sigma_{\rm{b}}^2 = \sigma_{\rm{[CII]}}^2 - \sigma_\mu^2$ & eq. \ref{eq:sigma_gr}\\
\hline
\end{tabular}

\end{table*}
\begin{figure*}
\centering
\includegraphics[height=0.95\textheight]{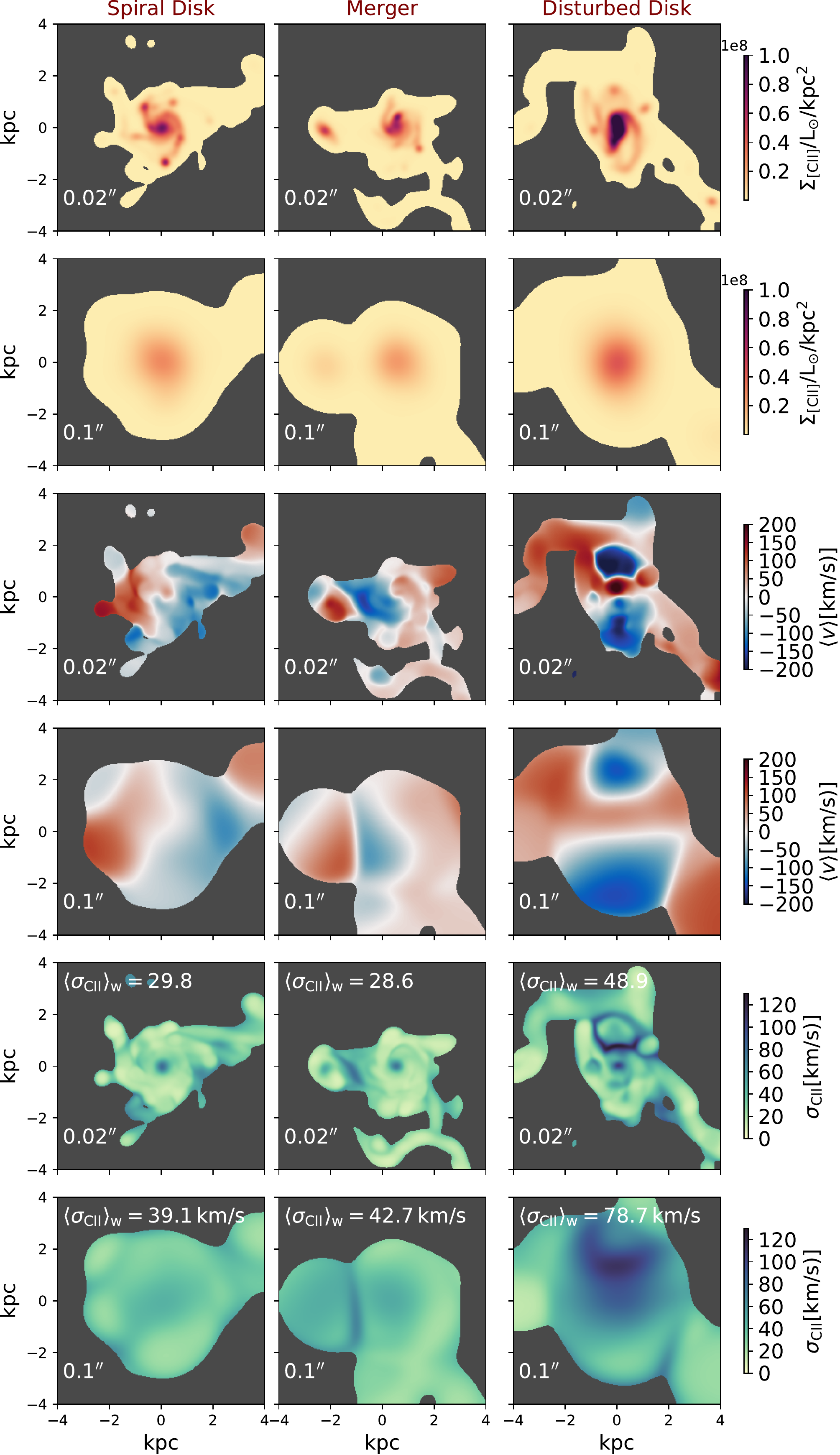}
\caption[]{Moment maps of evolutionary stages of Freesia with angular resolution $0.02^{\prime \prime}$ and $0.1^{\prime \prime}$. Surface luminosities are in linear scale. Notation as in Fig. \ref{fig:cii_moment_maps}.}
\label{fig:diff_res}
\end{figure*}

We focus our analysis on three different dynamical stages during the evolution of Freesia in the redshift range of $6<z<8$. The stages are denominated \quotes{\textit{Spiral Disk}} (SD), \quotes{\textit{Merger}} (MG), and \quotes{\textit{Disturbed Disk}} (DD).
In Tab. \ref{tab:stages_properties}, general properties of these stages are tabulated\footnote{Since the stages are selected from the evolution of a single galaxy, they have different stellar masses as well as SFR. Alternatively, one could look at different galaxies with the same stellar masses and SFR, but different dynamical structures. This will be considered in future work.}. Among the stages, the gas mass of Freesia ($M_g$) varies within a factor $\lsim 15\%$, while the star formation rate (SFR) and stellar masses $M_{\star}$ have a variation by a factor $\lsim 3$.
In particular, DD stage has the highest star formation rate ($\rm{SFR}\simeq 85.4\,\msolar\rm{yr}^{-1}$) and stellar mass ($M_\star \simeq 10^{10}\,\msolar$) while the MG stage has the lowest values ($\rm{SFR}\simeq 29.5\,\msolar\rm{yr}^{-1}$,  $M_\star \simeq 4\times10^{9}\,\msolar$ ). The total \CII~luminosity of all stages is similar and about $10^8 \rm{L}_\odot$, with the DD stage being the most luminous one. Overall, Freesia in these stages shows properties comparable to the bulk of the observed high-$z$ galaxies, reported in Table 1 by \citet{Kohandel+19}.

The typical radius in all stages is $r_d=1 \,{\rm kpc}$, thus the circular velocity can be estimated via
\be
v_{\rm c} = \sqrt{\frac{G M_{\rm{dyn}}}{r_d}}\,,
\ee
where $M_{\rm dyn}=M_{\rm g}+M_\star$ is the dynamical mass. Therefore, $v_{\rm c}= (189, 173, 246) \,\kms$ for the (SD, MG, DD) stages, respectively.

Note that -- similarly to \citet{Kohandel+19} -- different stages are identified and labeled based on the morphology of the \CII~line surface brightness maps and the corresponding (total) spectra extracted for their face-on and edge-on views, as can be appreciated from Fig. \ref{fig:cii_moment_maps}, where we show moment maps as well as the corresponding integrated spectra.

We start by looking at the moment-0 (surface brightness) maps for the face-on view\footnote{We orientate the l.o.s. parallel to the eigenvector of the inertia tensor of the gas density distribution with the largest eigenvalue.} of the three stages (second row of Fig. \ref{fig:cii_moment_maps}). Morphologically, the three stages are clearly discernible. The SD stage features a rotating disk with a one-sided, extended tail due to infalling gas; the MG stage is produced by a satellite merging into the main galaxy; the DD stage resembles a very complex structure as a consequence of the presence of a nearby, giant star-forming clump of gas (size of about $\sim 0.5\, \rm{kpc}$) perturbing the main galaxy disk. 

Rotating disks, mergers and disturbed disks have distinguishable spectral signatures in the \CII~spectra -- particularly for inclinations close to edge-on, even at very high redshifts \citep{Kohandel+19}. Looking at the edge-on spectra of the selected stages (first row of Fig. \ref{fig:cii_moment_maps}) we see that the SD stage shows a double-peak profile; instead, the signature of rotation in the spectra of the other two stages has been blurred by either the merging satellite (in MG stage) or the giant clump of gas hitting the disk (in DD stage). In each stage the reported value of $v_c$ (Tab. \ref{tab:stages_properties}) is roughly consistent with the half-width of the corresponding edge-on spectra, as expected from rotation support.

As discussed in Sec. \ref{sec:model_moments}, a key dynamical quantity obtained from HDCs is the spatially resolved mean velocity map, $\langle v \rangle$ (see eq. \ref{eq:mean_vel}), shown for the three stages in the third row of Fig. \ref{fig:cii_moment_maps}. The SD stage shows a well-formed velocity gradient in the central part of the system: this feature resembles a \quotes{spider diagram} pattern -- i.e. a well-known signature of rotating spiral galaxies \citep{Begeman1989} -- that is indicative of the existence of a rotating disk. The MG stage has two distinct rotating components, one for the main galaxy and the other for its satellite. The DD stage has a very complex velocity structure due to the presence of the giant clump disturbing the disk. These results show that a single galaxy might undergo dramatic changes in the course of its evolution, mostly arising from the complexity and intermittency of the assembly processes. As velocity dispersion encodes a record of the associated kinetic energy deposition, it provides a unique diagnostic tool to understand the build-up of these early systems.
\section{Characterising the velocity dispersion}\label{sec-dispersion}
\begin{figure*}
    \centering
    \includegraphics[width=\textwidth]{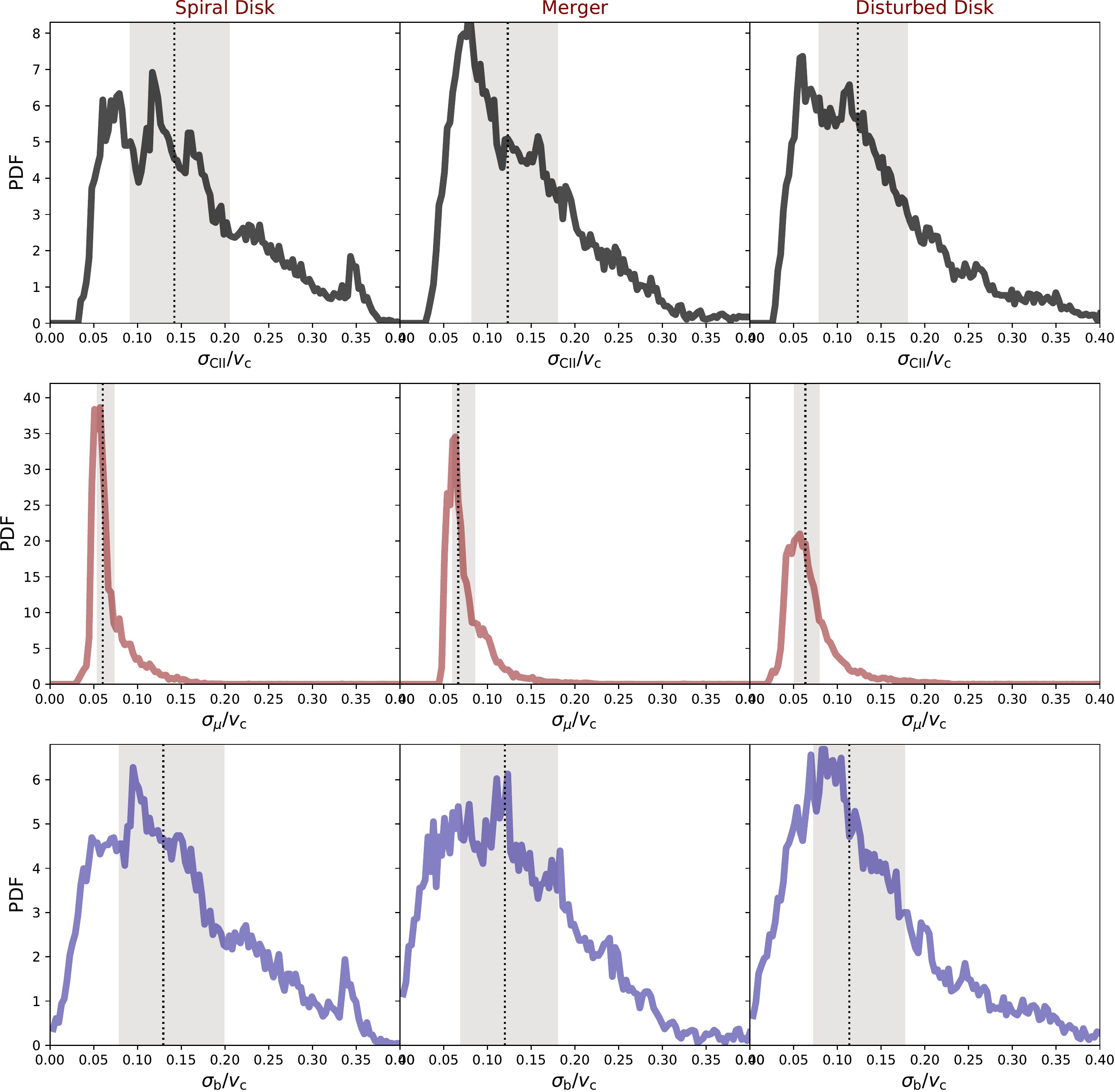}
    \caption{Probability distribution functions of the different velocity dispersion components for three evolutionary stages of Freesia.
    From top to bottom, we show the total velocity dispersion, $\sigma_{\rm [CII]} = (\sigma_{\mu}^2+\sigma_{b}^2)^{1/2}$,  from Fig. \ref{fig:cii_moment_maps}; small-scale velocity dispersion, $\sigma_{\mu}$, due to stellar feedback from Fig. \ref{fig:intrinsic-dispersion}; the large-scale velocity dispersion, $\sigma_{b}$, due to bulk motions from Fig. \ref{fig:sigma_gr}. See Tab. \ref{tab:sigma_list} for a summary of the definitions.
    The vertical dotted line denotes the mean value; the gray-shaded area represents the width of the distribution which is the difference between $25$th and $75$th  percentile of the distribution. 
    \label{fig:pdf_sigma}}
\end{figure*}

\begin{figure*}
\centering
\includegraphics[width=0.9\textwidth]{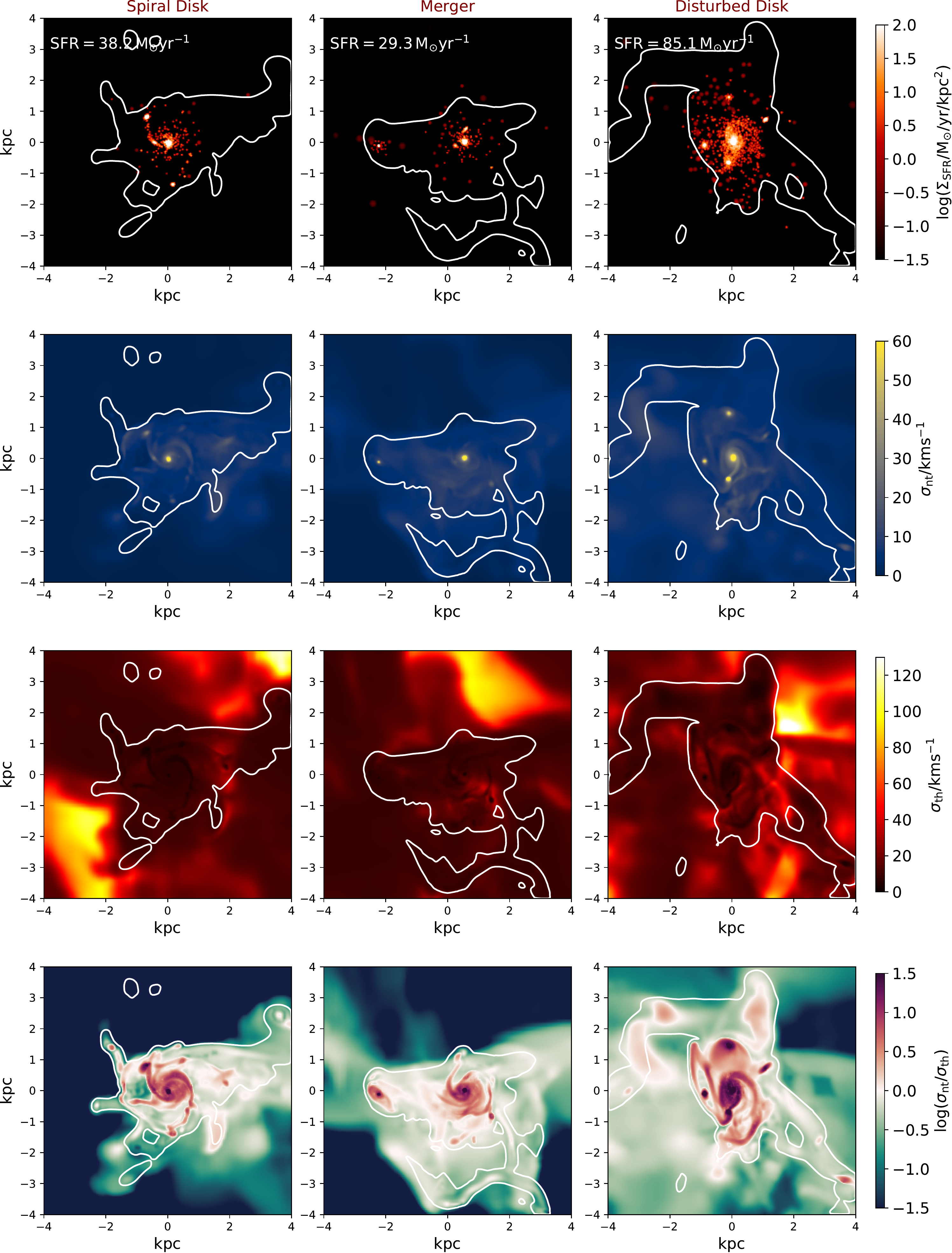}
\caption[Intrinsic line broadenings]{Star formation and stellar feedback in the different stages of Freesia.
{\it Top row}: star formation rate density maps {\it Second}: the velocity dispersion due to non-thermal pressure (turbulence), {\it Third}: thermal line broadening. {\it Bottom}: the ratio between non-thermal and thermal line broadening. White contours correspond to the $\Sigma_{\rm CII} = 10^{4.5}L_{\odot}\,\rm{kpc}^{-2}$ cuts of Fig. \ref{fig:cii_moment_maps}.
\label{fig:intrinsic-dispersion}
}
\end{figure*}

\begin{figure*}
\centering
\includegraphics[width=0.9\textwidth]{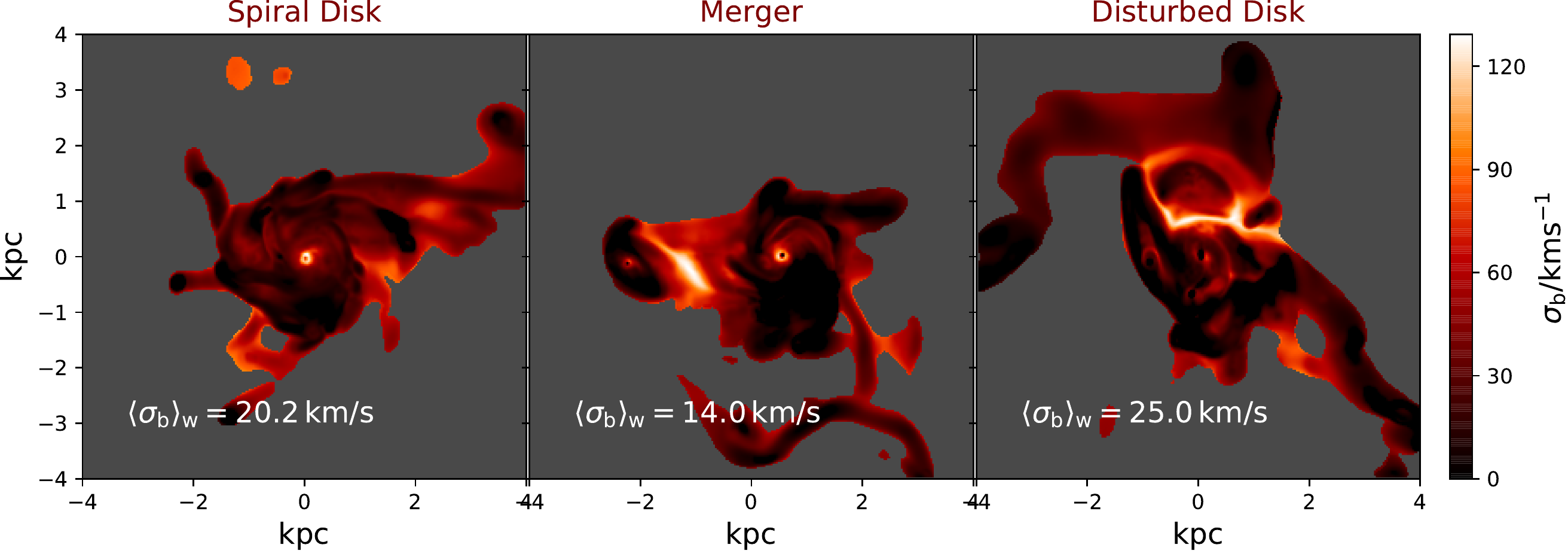}
\caption{Maps of velocity dispersion due to gravitational interactions (Eq. \ref{eq:sigma_gr}) for three stages of Freesia. Notation as in Fig. \ref{fig:cii_moment_maps}.
\label{fig:sigma_gr}
}
\end{figure*}

In this Section, we first compute the velocity dispersion by analyzing moment-2 maps of the \CII~line data cubes for the considered evolutionary stages (Sec. \ref{sec_tot_dispersion}). Then, in Sec. \ref{sec:drivers}, we assess the role of (i) stellar feedback and (ii) bulk motions in driving the velocity dispersion.

\subsection{Spatially resolved velocity dispersion maps}\label{sec_tot_dispersion}

In the fourth row of Fig. \ref{fig:cii_moment_maps}, we plot moment-2 ($\sigma_{\rm{CII}}$) maps for the three evolutionary stages of Freesia. On each plot, the \CII~luminosity-weighted average velocity dispersion\footnote{
$\langle \sigma_{\rm{CII}} \rangle_{\rm{w}} \equiv  {\sum{\sigma_{\rm{CII}}\Sigma_{\rm{[CII]}}}}/{\sum{\Sigma_{\rm{[CII]}}}}$}, i.e. $\langle \sigma_{\rm{CII}} \rangle_{\rm{w}}$ , is reported which reduces the $2$D maps to a single average value. As in the case of the $\langle v \rangle$ maps, the $\sigma_{\rm{CII}}$ maps are quite different, depending on the stage.
Nevertheless, there are common features. All the stages show a $\sigma_{\rm{CII}}$ peak up to $\sim 130\, \kms$ located at the galactic center; this is partially linked to the star formation activity, as we will see in Sec. \ref{sec:drivers}.

Apart from the central peak, the SD stage (with $\langle \sigma_{\rm{CII}} \rangle_{\rm{w}} = 25.7\, \kms$) has an almost uniform velocity dispersion map in the central $1\, \rm{kpc}$ region with a value of $\sim 15-20\,\kms$ with enhanced values ($\sigma_{\rm{CII}}$ up to $70\, \kms$) in the extended tail due to infalling gas.

The MG stage ($\langle \sigma_{\rm{CII}} \rangle_{\rm{w}} = 22.6\, \kms$) looks similar in most of the disk region, but a second peak (at $\sim 50\, \kms$) appears that corresponds to the center of the merging satellite. Moreover, $\sigma_{\rm{CII}}$ is boosted up to $\sim 80\, \kms$ as a ruslt of the bulk motions driven by the gravitational interaction between the main galaxy and the satellite (see Sec. \ref{sec:grav-turb}). 

Finally, the DD stage has the highest $\langle \sigma_{\rm{CII}} \rangle_{\rm{w}}= 36.5\, \kms$) values, and has a complex velocity dispersion structure paralleling that of the $\langle v \rangle$ map. There are various arcs in the central $1\, \rm{kpc}$ scale of the disk; the region, $1\,\rm{kpc}$ north of the center, features a pronounced disturbance likely due to the very close encounter of the giant clump of gas with the disk; such very high $\sigma_{\rm{CII}}$ (up to $\sim 130\,\kms$), the elongated region extends for about $1\,\rm{kpc}$ following the circumference of the disk. Such a feature in velocity dispersion maps is due to gravitational interactions of multi-component systems (in this case, the main galaxy and the giant gas clump), as we detail in Sec. \ref{sec:grav-turb}. 

To summarize, if we measure the level of ISM velocity dispersion via a luminosity-weighted average value as i.e. done in actual observations (like e.g. \citealt{Green+14}), Freesia shows a moderate value around $\sim 23-38\, \kms$. Note that these values are obtained for an angular resolution of $\sim 0.0005^{\prime \prime}$, which however impacts the conclusions, as we will see in Sec. \ref{sec:beam-smaring}.

Using the galaxy circular velocity and the average l.o.s velocity dispersion, we can define the rotational-to-dispersion support ratio, $v_{\rm c}/\sigma$. Adopting $\sigma = \langle \sigma_{\rm{CII}} \rangle_{\rm{w}}$ leads to a  $v_{\rm c}/\sigma$ ratio of 7.4, 7.7 and 6.7 for the cold \CII~emitting gas in SD, MG and DD stage respectively. As a comparison, such ratio is $\approx 20$ for the MW, and $3.5-6$ for intermediate redshift galaxies \citep{Hodge+12,Swinbank+11}. Thus, cold gas in EoR galaxies -- unlike the MW but similarly to galaxies at comic noon -- receives considerable support from random motions.

\subsection{Beam smearing effects}\label{sec:beam-smaring}
We want to understand the effect of beam smearing on the resultant
dynamical observables derived from the full resolution ($0.0005^{\prime \prime}$) \CII~line HDC. 
We mimic the beam smearing by performing a smoothing with a Gaussian kernel of $0.02$ and $0.1$ arcsecs. Then using eqs. \ref{eq:moments}, we obtain the low-resolution counterparts of Fig. \ref{fig:cii_moment_maps}.
In Fig. \ref{fig:diff_res}, we show different moment maps of \CII~line for two angular resolutions ($0.02$ and $0.1$ arcsecs) for the evolutionary stages of Freesia. 

As expected, the beam smearing affects the morphology of various moment maps. In the lowest resolution case, it is very challenging to derive morphological/structural properties like the presence of a disk, satellites or clumps in each structure. This issue is very important and needs to be studied but it is beyond the scope of this paper.

As shown in Fig. \ref{fig:diff_res}, $\langle \sigma_{\rm{CII}} \rangle_{\rm{w}}$ \textit{increases with decreasing angular resolution}. More precisely, at $0.02$ arcsecs, it ranges between $28-50\,\kms$ while at $0.1$ arcsecs, it rises up to  $40-80\,\kms$. This effect is more dramatic in the DD stage. With the lowest resolution, $\langle \sigma_{\rm{CII}} \rangle_{\rm{w}}$ is doubled compared to the high-resolution case (Fig. \ref{fig:cii_moment_maps}). This effect is more severe in the DD stage because, as we saw in Sec. \ref{sec_tot_dispersion}, in the central $\sim 2\,\rm{kpc}$ part of the galaxy there are various arcs with large velocity dispersions $\simlt 130\,\kms$, substantially contributing to \CII~emission. When we perform the smoothing, the emission from these high dispersion arcs spreads over the disk (see the $\sigma_{\rm{CII}}$ DD map at  $0.1$ arcsec resolution). This yields a very large average velocity dispersion.

These results show that, when dealing with real observations, the beam smearing effect must be carefully accounted for when inferring the proper velocity dispersion of the system, particularly the interacting ones. For the rest of the paper, we continue our analysis with the high-resolution cases (Fig. \ref{fig:cii_moment_maps}), unless otherwise stated.

\section{Physical drivers}\label{sec:drivers}

To further investigate the structure of the observed velocity dispersion, it is convenient to build and analyze its $1$-D probability distribution function (PDF).
In the first row of Fig. \ref{fig:pdf_sigma}, we plot the (normalized) PDFs of the distribution of $\sigma_{\rm{CII}}/v_{\rm c}$ for three evolutionary stages of Freesia. The three PDFs have similar shapes: they peak at $\sigma_{\rm{CII}}/v_{\rm c} < 0.1$ and have a high velocity dispersion tail. All the distributions have a similar width, i.e. the difference between $25$th and $75$th percentile, around $0.1$.  Regarding the shape of the distribution, while the MG stage shows a single sharp peak at the low $\sigma_{\rm{CII}}$ part of the distribution, the SD and DD stages have a multiple peak structure. In particular, the SD stage has an additional peak in the high velocity dispersion  ($\sigma_{\rm{CII}}/v_{\rm c}\sim 0.35$) part of the distribution. To understand the physical origin of the main features in the PDF, it is necessary to quantify the individual contribution from different driving mechanisms to the observed velocity dispersion.

To identify the physical drivers of the observed velocity dispersion, $\sigma_{\rm{CII}}$, we start by investigating the effect of stellar feedback as a driver of velocity dispersion. Next, we turn to bulk motions sourced by gravitational interactions. 

\subsection{Stellar feedback}\label{sec:micro-turb}
In Fig. \ref{fig:intrinsic-dispersion} (first row), we plot Freesia's star formation rate density maps (eq. \ref{eq:sfr}).
For all of the stages, the SFR density is ($\Sigma_{\rm SFR}>10^{2}\,\msolar{\rm{yr}^{-1}}\rm{kpc}^{-2}$) at the galaxy center. The SD stage with a total SFR of $38.2\,\msolar \rm{yr}^{-1}$ has the smoothest $\Sigma_{\rm SFR}$ map, with most of the star formation occurring in the galactic disk. Interestingly, some bright star-forming regions are located along the spiral arms. The MG stage has the lowest total SFR ($29.3\,\msolar \rm{yr}^{-1}$). Signs of recent star formation are seen both on the main galaxy and the satellite. The DD stage has the highest total SFR ($85.1\,\msolar \rm{yr}^{-1}$), with various very bright star-forming sites along the spiral arms, as well as in the giant gas clump. 

In Fig. \ref{fig:intrinsic-dispersion}, we plot the turbulent, i.e. $\sigma_{\rm{nt}}$, (second row), and thermal velocity dispersion, i.e. $\sigma_{\rm{th}}$, maps (third row) induced by stellar feedback for the evolutionary stages of Freesia. On these maps, we have overplotted contours inside which pixels have $\Sigma_{\rm CII} > 10^{4.5}L_{\odot}\rm{kpc}^{-2}$ (the same luminosity cut used in the mean velocity and velocity dispersion maps in Fig. \ref{fig:cii_moment_maps}). In Tab. \ref{tab:stat-dispersion}, the \CII~luminosity weighted average values of the turbulent, $\langle \sigma_{\rm{nt}} \rangle_{\rm{w}}$, and thermal $\langle \sigma_{\rm{th}} \rangle_{\rm{w}}$ velocity dispersion are tabulated for the three evolutionary stages.

The $\sigma_{\rm{nt}}$ maps are almost flat but show high values at the galactic center ($\sigma_{\rm{nt}} = 100-140\,\kms$ depending on the stage). The SD and MG stages have similar $\langle \sigma_{\rm{nt}} \rangle_{\rm{w}} \simeq 11\, \kms$, while the DD stage has a slightly higher value ($\sim 18\, \kms$). This behaviour was expected since the SD and MG stage have similar SFRs (see Tab. \ref{tab:stages_properties}) while the MG stage has $\approx 3 \times$ higher SFR ($\sim 85 \msolar \rm{yr}^{-1}$); it also has more star-forming sites with high SFR density compared to the other stages (see Fig. \ref{fig:intrinsic-dispersion}). Thus, the higher the star formation, the more turbulent the ISM becomes due to the collective kinetic energy deposition by SNe, stellar winds and radiation pressure.

Instead, looking at $\sigma_{\rm{th}}$ maps, we see some shock-heated extended regions with $\sigma_{\rm{th}}>80\, \kms$ in addition to smooth central ($<1\, \rm{kpc}$) parts with low dispersion values ($\sim 10\,\kms$). Most of the shock-heated regions (for instance the south-west corner of the SD or a triangular region in the north of the MG stage map) are regions with very low \CII~line intensity ($\Sigma_{\rm{[CII]}} < 10^{4.5}\, \rm{L}_{\odot}/\rm{kpc}^2$). In terms of average values, all the stages have similar $\langle \sigma_{\rm{th}} \rangle_{\rm{w}}$ ($\sim 10\, \kms$; see Tab. \ref{tab:stat-dispersion}). 

In the bottom row of Fig. \ref{fig:intrinsic-dispersion}, we plot the ratio between the turbulent and thermal velocity dispersion as a proxy of the level of turbulence. The structure of the small-scale turbulence of the ISM of Freesia can be divided in three phases; sub-sonic ($\mathrm{log}(\sigma_{\mathrm{nt}}/\sigma_{\mathrm{th}})\le0$), supersonic ($0<\mathrm{log}(\sigma_{\mathrm{nt}}/\sigma_{\mathrm{th}})<1.5$) and hyper-sonic ($\mathrm{log}(\sigma_{\mathrm{nt}}/\sigma_{\mathrm{th}})\ge1.5$). In Freesia, the turbulence in most of the \CII~emitting gas is either supersonic (or even hypersonic). 
To quantify the total contribution of stellar feedback in $\sigma_{\rm{CII}}$,
we introduce the small-scale velocity dispersion \footnote{Note that to compute the small-scale velocity dispersion we have just accounted for the pixels with $\Sigma_{\rm{[CII]}} > 10^{4.5}\, \rm{L}_{\odot}/\rm{kpc}^2$} ($\sigma_\mu$) as 
\be\label{eq:sigma_mu_def}
\sigma_{\mu} = \sqrt{\sigma_{\rm{th}}^2 + \sigma_{\rm{nt}}^2 }\,.
\ee

In the middle row of Fig. \ref{fig:pdf_sigma}, we show the PDFs of $\sigma_{\mu}/v_{\rm c}$. For all stages, the distribution can be fitted with a single Gaussian function apart from the tail of the distribution specially in the SD and MG stage. The excess in the high velocity dispersion tail of the distribution in these stages is due to pixels with hyper-sonic turbulence for which $\mathrm{log}(\sigma_{\mathrm{nt}}/\sigma_{\mathrm{th}})\ge 1.5$ (see Fig. \ref{fig:intrinsic-dispersion}]). These high values of velocity dispersion will dissipate on time scales of $\lsim 0.1 \rm Myr$ (see eq. \ref{eq:dissipation}). 
The PDFs of $\sigma_{\mu}/v_{\rm c}$ with respect to $\sigma_{\rm{CII}}/v_{\rm c}$ (see first row of Fig. \ref{fig:pdf_sigma}) are very narrow and confined. Most of the distribution of $\sigma_{\rm{CII}}/v_{\rm c}$ can not be fully described by $\sigma_{\mu}$ only. Therefore, stellar feedback alone is not sufficient to maintain the observed $\sigma_{\rm{CII}}$; bulk motions arising from gravitational forces are then required.

\begin{table}
\caption{\CII~luminosity weighted velocity dispersion values.\label{tab:stat-dispersion}}
\begin{center}
\begin{tabular}{|l|c|c|c|c|}
\hline
Stages &$\langle \sigma_{\rm{CII}} \rangle_{\rm{w}}$& $\langle \sigma_{\rm{nt}} \rangle_{\rm{w}}$&  $\langle \sigma_{\rm{th}}\rangle_{\rm{w}}$& $\langle  \sigma_{\rm{b}}\rangle_{\rm{w}}$\\
~                   & $[\kms]$ & $[\kms]$ & $[\kms]$ & $[\kms]$ \\
\hline
Spiral Disk (SD)    & $25.7$   & $11.5$   & $7.0$    & $20.7$   \\
Merger (MG)         & $22.6$   & $11.8$   & $10.6$   & $17.6$   \\
Disturbed Disk (DD) & $36.5$   & $18.4$   & $10.5$   & $27.8$   \\
\hline
\end{tabular}
\end{center}
\end{table}

\subsection{Bulk motions}\label{sec:grav-turb}

We define the bulk velocity dispersion as 
\be\label{eq:sigma_gr}
\sigma_{\rm{b}} = \sqrt{\sigma_{\rm{[CII]}}^2 - \sigma_{\mu}^2}\,.
\ee
One of the obvious sources of bulk motions is the rotational energy of the system. This can be directly subtracted out from the analysis if we consider, as done here, the face-on view of the galaxy. Other than rotation, the velocity dispersion can be increased by disordered, large-scale motions generated by gravitational interactions occurring in multi-component systems like Freesia. 

For instance, the velocity enhancement mentioned earlier, and due to infalling gas (in the SD stage), merging satellites (MG), or close encounters with clumps (DD), fall in this category (see Fig. \ref{fig:cii_moment_maps}). These enhancements cannot be explained by the stellar feedback that we analyzed in the previous Section. In Fig. \ref{fig:sigma_gr}, we show  $\sigma_{\rm{b}}$ maps for the three evolutionary stages of Freesia. In terms of average values, the DD stage has the highest $\langle \sigma_{\rm{b}} \rangle_{\rm{w}}\simeq 28\, \kms$ value, whereas the MG stage only reaches $\simeq 18\, \kms$. We also plot the $\sigma_{\rm{b}}/v_{\rm c}$ PDF in the bottom row of Fig. \ref{fig:pdf_sigma}. By comparing the PDF of $\sigma_{\rm{CII}}/v_{\rm c}$ and $\sigma_{\rm{b}}/v_{\rm c}$, we conclude that the high velocity tail of the $\sigma_{\rm{CII}}$ distribution is largely produced by bulk motions such as gravitational interactions. Hence, gravity  provides the dominant  ($> 90\%$) contribution to kinetic energy observed in \CII . 

A key difference between bulk motions and feedback-related turbulence is the dissipation time. Referring to eq. \ref{eq:dissipation}, the dissipation time scale for $\sigma_{\rm b}$ would be of the order of $l_{\rm b}/ \sigma_{b}$, where $l_{\rm b}$ can be conservatively taken as the disk radius, or  $l_b \approx 1 {\rm kpc} = 100\, l_{\rm cell}$. Hence, the dissipation time of bulk motions is about 30-50 times longer than that of small-scale turbulence produced by energy injection from massive stars. In turn, this allows gravitationally-induced motions to dominate the overall kinetic energy budget of the galaxy.

\section{Spatially resolved $\sigma_{\rm CII} - \Sigma_{\rm SFR}$ relation}\label{sec-sigma-sfr}

\begin{figure*}
\centering
\includegraphics[width=\textwidth]{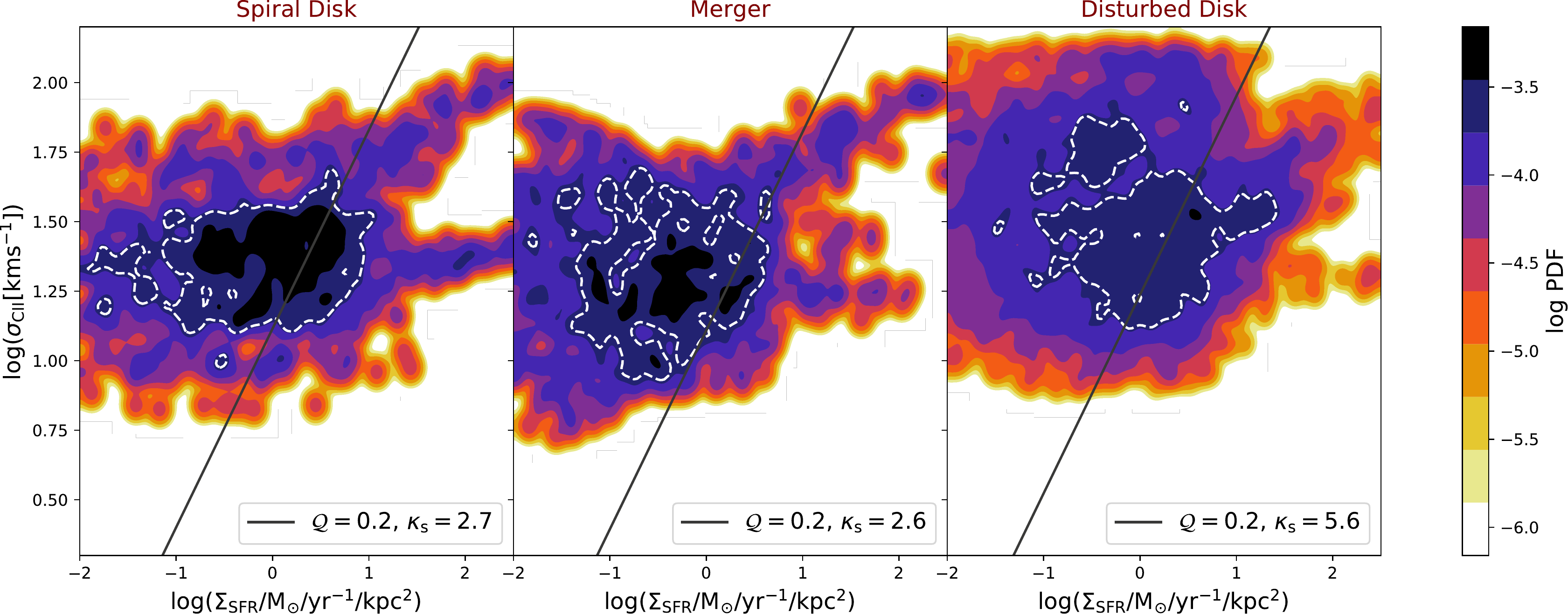}
\caption{2D PDFs of spatially resolved l.o.s velocity dispersion derived from \CII~data cubes ($\sigma_{\mathrm{CII}}$) and star formation rate densities for various stages of Freesia. The white dashed contours show the region including $90\%$ of the data. The solid lines indicate the analytical expression for $\sigma-\Sigma_{\rm{SFR}}$ relation (see eq. \ref{eq:sigma_SigmaSFR} and values given in Tab. \ref{tab:stages_properties}).
\label{fig:sigma-sfr-2d}
}
\end{figure*}

\begin{figure*}
\centering
\includegraphics[width=\textwidth]{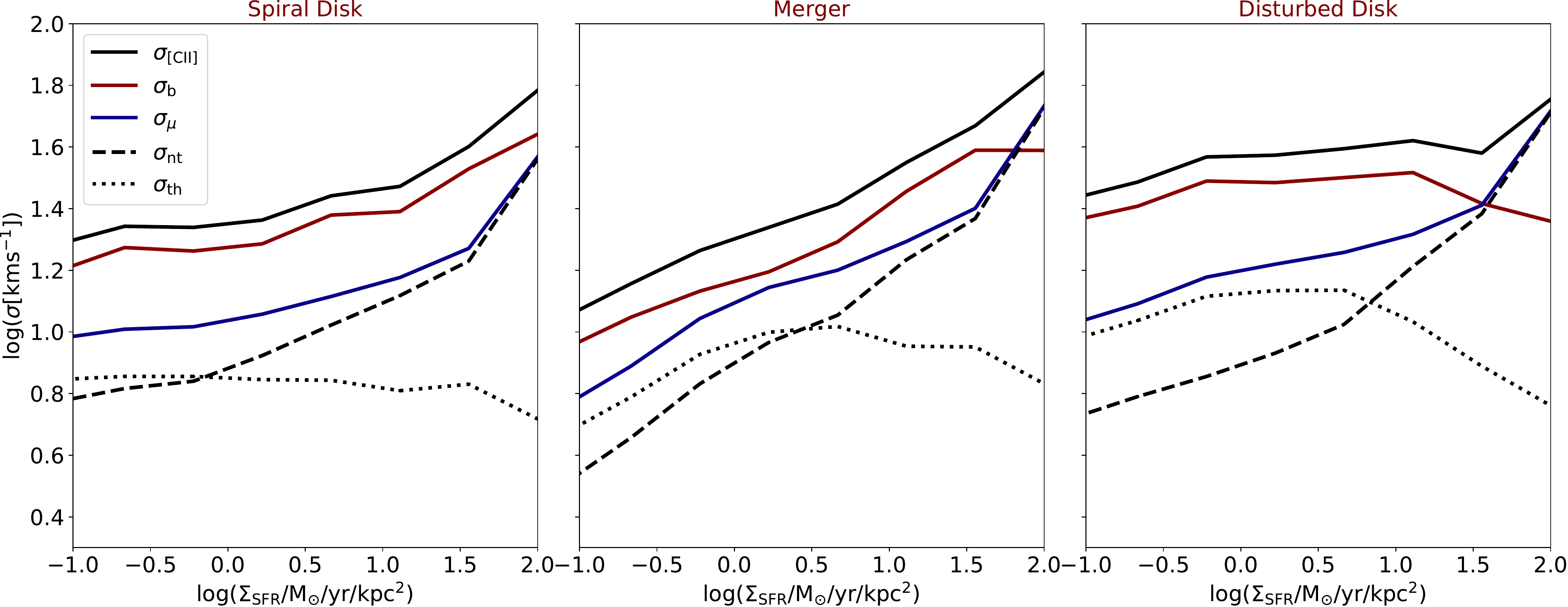}
\caption{The average relation between different components (see internal label) of the total velocity dispersion, $\sigma_{\rm CII}$, and the star formation surface density $\Sigma_{\rm SFR}$. Each panel refers to a different evolutionary stage. The curves are obtained from the data shown in Fig. \ref{fig:sigma-sfr-2d} by averaging the data in $10$ bins of SFR density in the range $0.1-100\,\msun\rm{yr}^{-1}\rm{\rm kpc}^{-2}$. 
\label{fig:sigma-sfr} 
}
\end{figure*}

Using the results in Sec. \ref{sec:drivers}, we can study the local relation between the l.o.s velocity dispersion and star formation rate densities in Freesia. We compute the 2D distribution of different velocity dispersion components ($\sigma_{\rm CII},\sigma_{\rm nt},\sigma_{\mu}, \sigma_b, \sigma_{\rm CII}$) as a function of $\Sigma_{\rm SFR}$. In Fig. \ref{fig:sigma-sfr-2d}, we concentrate on the distribution of total $\sigma_{\rm{CII}}$ for the SD, MG and DD stage. We do not see a clear correlation between these quantities. It seems that $\sigma_{\rm{CII}}$ for all the stages is constant for a range of $4$ dex of star formation rate densities.

To see the behaviour of other components of the velocity dispersion ($\sigma_{\rm CII},\sigma_{\rm nt},\sigma_{\mu}, \sigma_b, \sigma_{\rm CII}$), we obtain the average $\sigma$-$\Sigma_{\rm SFR}$ relation for each component by averaging the data in $10$ bins of SFR density in the range $0.1-100\,\msun\rm{yr}^{-1}\rm{\rm kpc}^{-2}$. The results are shown in Fig. \ref{fig:sigma_gr} for the usual three evolutionary stages.

First we concentrate on $\sigma_{\rm{CII}}$. For the SD and DD stages, $\sigma_{\rm{CII}}$ is almost independent on $\Sigma_{\rm SFR}$ over three orders of magnitude, apart from a slight increase (factor $\lsim 1.5$) at the high-end of the star formation range.
Such a result is in agreement with a recent theoretical work \citet{Orr+19}. These authors study the relation between gas velocity dispersion and star formation rate for Milky Way-like galaxies in \code{FIRE-2} simulations \citep{Hopkins+18}. They find a relatively flat relation ($\sigma \sim 15-30\, \kms$ in neutral gas) across 3 dexes in SFR; this is also in agreement with nearby galaxies observations \citep{Zhou+17}. Note that \citet{Orr+19} do not model emission lines to compute the l.o.s velocity dispersion. This might affect their conclusion as they are not directly computing the observed velocity dispersion, which can be affected by resolution issues (see Fig. \ref{fig:diff_res}).

For the MG stage, the situation is different. The dependence of $\sigma_{\rm{CII}}$ on $\Sigma_{\rm SFR}$ shows instead an increasing trend across the SFR range. This correlation might be partly caused by the fortuitous presence of a satellite with peculiar properties; The satellite has low $\sigma_{\rm{CII}}$, low SFR. Because of its large \CII~emission, it dominates the low SFR part of the relation decreasing the \CII-weighted velocity dispersion and therefore $\sigma_{\rm{CII}}$ is biased-low.

The different contributions to $\sigma_{\rm{CII}}$ follow similar trends with star formation in the three stages: $\sigma_{\rm{b}}$ always dominates the relation for $\Sigma_{\rm SFR} \simlt 10-30\, \,\msun\rm{yr}^{-1}\rm{\rm kpc}^{-2}$, i.e. bulk motions such as gravitational interactions are the main drivers of the velocity dispersion in moderate star-forming, high-redshift galaxies.
At higher SFRs, stellar feedback becomes important and $\sigma_{\mu}$ catches up with bulk motions. The increase of $\sigma_{\mu}$ at high SFR is due to enhanced momentum injection by massive stars powering supersonic turbulence for which $\sigma_{\mathrm{nt}}/\sigma_{\mathrm{th}}>3$. The contribution from supersonic regions to the total \CII~luminosity is $\approx 10\%$ for the SD and DD stages and $\approx 5\%$ for the MG stage. The thermal component, $\sigma_{\rm{th}}$ becomes larger than the turbulent term ($\sigma_{\rm{nt}}$) only at $\Sigma_{\rm SFR} \lsim 5\, \,\msun\rm{yr}^{-1}\rm{\rm kpc}^{-2}$. 
In general, the feedback-related turbulent level in Freesia achieves an almost constant value of $\sigma_\mu \simeq 10-15\, \kms$ independently of SFR. Such value is higher than the velocity dispersion typically observed in local molecular clouds, which is around $5-10\,\kms$ \citep{Bolatto+2008}. In summary, the ISM velocity dispersion in assembling, EoR galaxies appears to be dominated by the bulk motions component produced by gravitational interactions, such as accretion/merging events.

To gain some insight, it is instructive to compare our results with a simple physical model for the $\sigma - \Sigma_{\rm SFR}$ relation. The Toomre parameter, $\mathcal{Q}$ \citep{Toomre+1964} for a galaxy with a total mass surface density $\Sigma = \Sigma_{\rm{g}} + \Sigma_\star$, l.o.s velocity dispersion $\sigma$, and epicyclic frequency $\kappa={a v_{\rm{c}}}/{r_d}$, where $a=\sqrt{2}$ for galaxies with flat rotation curves \citep{inoue:2016_toomre,leung:2019} is given by
\be \label{eq:toomre}
\mathcal{Q} = \frac{\sigma \kappa}{\pi G \Sigma} = \sqrt{2}\left( \frac{v_{\rm c}}{\sigma} \right)^{-1}
\ee
where $G$ is the gravitational constant.
The total mass surface density is computed from the data in Tab. \ref{tab:stages_properties} assuming as a reference radius, $r_d=1$ kpc, for both gas and stars. 

We introduce the gas fraction, $f_g = \Sigma_g /(\Sigma_\star +\Sigma_{g})$; this is given in Tab. \ref{tab:stages_properties} for Freesia. Then assuming an average Toomre parameter $\mathcal{Q}$ for the galaxy, we can relate the l.o.s velocity dispersion to the gas surface density as
\be
\sigma = \mathcal{Q} \sqrt{\frac{\pi {G} \Sigma_{g} r_d}{ 2 f_g}}.
\ee

To relate the gas surface densities to star formation rate surface densities, we assume a generalized KS relation \citep{Heiderman+10,pallottini:2019,ferrara:2019}
\be\label{eq_ks}
\left ( \frac{\Sigma_{\rm{SFR}}}{\msolar \rm{yr}^{-1}\rm{kpc}^{-2}}\right ) = 10^{-12} \kappa_{\rm{s}}\left( \frac{\Sigma_{\rm{g}}}{\msolar \rm{kpc}^{-2}}\right )^{n}\,,
\ee
with $n=1.4$; $\kappa_{\rm{s}}$ is the burstiness parameter, given in Tab. \ref{tab:stages_properties}, expressing deviations from the empirical local relation \citep{Kennicutt+98}. For starburst galaxies, $k_s > 1$.
Hence, the relation between gas l.o.s velocity dispersion $\sigma$ and star formation rate surface density $\Sigma_{\rm{SFR}}$ is: 
\begin{subequations}\label{eq:sigma_SigmaSFR}
\be 
{\sigma} = 70\, A\left ( \frac{\Sigma_{\rm{SFR}}}{\msolar \rm{yr}^{-1}\rm{kpc}^{-2}} \right )^{5/7}\left( \frac{r_d}{\rm{kpc}}\right)^{1/2}\,\kms,
\ee
where 
\be
A=\frac{1}{\sqrt{2}}\frac{\mathcal{Q}}{  f_g^{1/2} k_s^{5/7}}\,
\ee
\end{subequations}
which requires the information on the average $\mathcal{Q}$ parameter.

We calculate the average Toomre parameter for \CII~emitting gas in three stages of Freesia by using eq. \ref{eq:toomre} and $\sigma = \langle \sigma_{\rm CII}\rangle_{\rm w}$. We derive $\mathcal{Q}\simeq 0.2$ for all the three stages, a value compatible, but slighlty lower, than typically deduced for intermediate redshift galaxies ($\mathcal{Q}\sim 0.5$ in \citealt{Swinbank+11} and $\mathcal{Q}\sim 0.25$ in \citealt{Hodge+12}). 
Note that all the stages are in a starburst phase, with $\kappa_{\rm s}$ between 2.6 and 5.6, in an agreement with what inferred by \citet{vallini:2020} for COS-3018 -- a $z\simeq 6.8$ redshift galaxy -- from UV, CIII] and [CII] data \citep{Carniani+18himiko,smith2018Natur,laporte:2017cos}.

In Fig. \ref{fig:sigma-sfr-2d}, we overplot Eq. \ref{eq:sigma_SigmaSFR} on top of simulations data for the three stages of Freesia. Although the simulated PDF of $\sigma_{\rm CII}-\Sigma_{\rm SFR}$ does not show a clear trend, $90\%$ data inclusion regions lie on the derived average analytical expression for the SD and DD stages. The actual velocity dispersion in the MG stage is instead higher than predicted by the analytical relation; this is expected given the simplifying, thin disk assumptions on which the latter is based.

Although we have concluded that stellar feedback plays a sub-dominant role in determining the observed velocity dispersion, we should warn that the delay between star formation and the corresponding feedback can introduce complications in this picture \citep{Orr+19a}. In addition, the delay depends on the specific feedback process considered. For example, the delay time is $5-30\, \rm{Myr}$ for supernova feedback, and $0-10\, \rm{Myr}$ for ionizing radiation and winds from OB stars \citep{Leitherer+99}. However, the observed $\sigma-\Sigma_{\rm SFR}$ relation \citep[][]{Lehnert+13,Yue+19}) is obtained from star formation rates and the velocity dispersions measured at the same time. We have roughly accounted for this bias by including only stars younger than $30$ Myr in the simulated SFR computation (see Sec. \ref{sec:feedback}).

\section{Summary}

We have studied the structure of the spatially resolved line of sight velocity dispersion for galaxies in the EoR traced by \CII~line emission. Our laboratory is a galaxy in the \code{SERRA} suite of zoom-in simulations called \quotes{Freesia}. 

We have modelled \CII~emission Hyperspectral Data Cubes (HDC) for three evolutionary stages of Freesia: \textit{Spiral Disk} (SD) at $z=7.4$, \textit{Merger} (MG) ($z=8.0$), and \textit{Disturbed Disk} (DD) ($z=6.5$). These three stages correspond to well-defined, distinct dynamical states of the galaxy. SD is a rotating disk ($v_{\mathrm{c}}=189\, \kms$) with an extended, lopsided tail due to the in falling gas; the MG stage (with $v_{\mathrm{c}}=173\, \kms$) contains a satellite merging into the main galaxy; finally, the DD stage ($v_{\mathrm{c}}=246\, \kms$) is the most complex configuration due to the presence of a giant clump of gas (size about $\sim 0.5\, \rm{kpc}$) in the very vicinity of the main galaxy. The total \CII~luminosity of the stages is $\simeq 10^8\, L_{\odot}$.
From the simulated HDC, we have built spatially resolved mean velocity $\langle v \rangle$, and velocity dispersion $\sigma_{\rm{CII}}$ maps for all the stages. Using $\sigma_{\rm{CII}}$ maps we have evaluated the level of velocity dispersion in the ISM of high-$z$ galaxies and determined its physical drivers. We have studied the contribution of velocity dispersion due to bulk motions ($\sigma_{\rm{b}}$) and stellar feedback ($\sigma_\mu$), the latter incorporating both turbulent ($\sigma_{\rm{nt}}$) and thermal ($\sigma_{\rm{th}}$) small-scale contributions. Finally, we have investigated the existence of a relationship between different components of velocity dispersion and the star formation rate. The main results of this work can be summarized as follows:

\begin{itemize}
  \item[$\bullet$] We have quantified the \CII~luminosity-weighted average velocity dispersion $\langle \sigma_{\rm{CII}} \rangle_{\rm{w}}$. At the full resolution of our simulation, which is equivalent to an angular resolution of $0.005^{\prime \prime}$, we find $\langle \sigma_{\rm{CII}} \rangle_{\rm{w}}= (25.7, 22.6, 36.5)\, \kms$ for the (SD, MG, DD) stages, respectively. Hence, we conclude that Freesia has a moderate average velocity dispersion regardless of the stage.

  \item[$\bullet$] $\langle \sigma_{\rm{CII}} \rangle_{\rm{w}}$ is very sensitive to the angular resolution of the observations. Due to beam smearing effects, the average value increases at lower resolutions. This effect is more severe for actively interacting systems, exemplified by our DD stage. Observations with an angular resolution of $0.02^{\prime \prime}$ ($0.1^{\prime \prime}$), would infer an average velocity dispersion $16-34\%$ ($52-115\%$) larger than the actual one.

  \item[$\bullet$] We have calculated the rotational-to-dispersion support ratio as well as Toomre $\mathcal{Q}$ parameter using $\langle \sigma_{\rm{CII}} \rangle_{\rm{w}}$ for Freesia. We derive $v_{\mathrm{c}}/\sigma \simeq 7$ and $\mathcal{Q}\simeq 0.2$ suggesting that \CII~emitting cold gas in EoR galaxies -- unlike the MW but similar to galaxies at cosmic noon -- receives considerable support from random motions.

  \item[$\bullet$] Concerning the resolved $\sigma_{\rm{CII}} - \Sigma_{\rm SFR}$ relation, we find  a relatively flat relation for $0.02< \Sigma_{\rm SFR}/\,\msun\rm{yr}^{-1}\rm{\rm kpc}^{-2} < 30$. The majority of simulated data lies on the derived average analytical expression, i.e. $\sigma \propto \Sigma_{\rm SFR}^{5/7}$ for the SD and DD stages. However, in the MG stage, the actual velocity dispersion is somewhat higher than predicted by the analytical expression, due to the simplifying assumptions on which the latter is based.

  \item[$\bullet$] Stellar feedback yields a $\sigma_{\mu}\simeq 10-15\, \kms$ almost independently from the total SFR, due to the balance between energy injection by massive stars, and the rapid dissipation of small-scale supersonic turbulence. However, the stellar feedback accounts only for $< 10\%$ of the total kinetic energy. We conclude that at high-redshift the velocity dispersion is dominated by bulk motions produced by gravitational interactions -- such as accretion/merging events, that govern the build-up phase of EoR galaxies.
  
\end{itemize}

\section*{Acknowledgements}
MK, AF, and SC acknowledge support from the ERC Advanced Grant INTERSTELLAR H2020/740120.
LV acknowledges funding from the European Union's Horizon 2020 research and innovation program under the Marie Sklodowska-Curie Grant agreement No. 746119.
We acknowledge use of the Python programming language \citep{VanRossum1991}, Astropy \citep{astropy}, Cython \citep{behnel2010cython}, Matplotlib \citep{Hunter2007}, NumPy \citep{VanDerWalt2011}, \code{Pymses} \citep{Labadens2012}, \code{pynbody} \citep{pynbody}, and SciPy \citep{scipy2019}.
\section*{Data Availability}
The data underlying this article were accessed from the computational resources available to the Cosmology Group at Scuola Normale Superiore, Pisa (IT). The derived data generated in this research will be shared on reasonable request to the corresponding author.


\bibliographystyle{mnras}
\bibliography{master,codes}     

\appendix
\section{Small-scale velocity dispersion}

\begin{figure*}
\centering
\includegraphics[width=\textwidth]{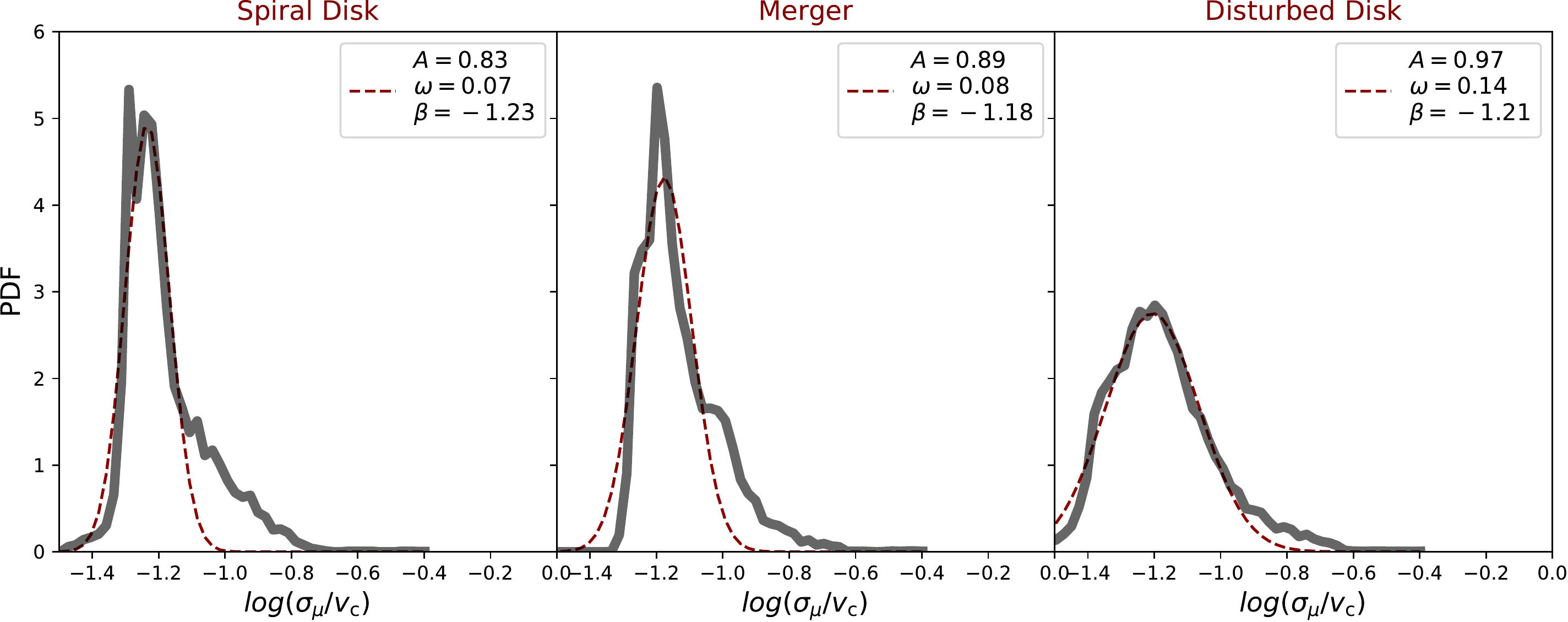}
\caption{Probability distribution function of small-scale velocity dispersion for three stages of Freesia along with the fitted normal distribution (dashed line).
\label{fig:sigma-mu-pdf}
}
\end{figure*}
\begin{figure*}
\centering
\includegraphics[width=\textwidth]{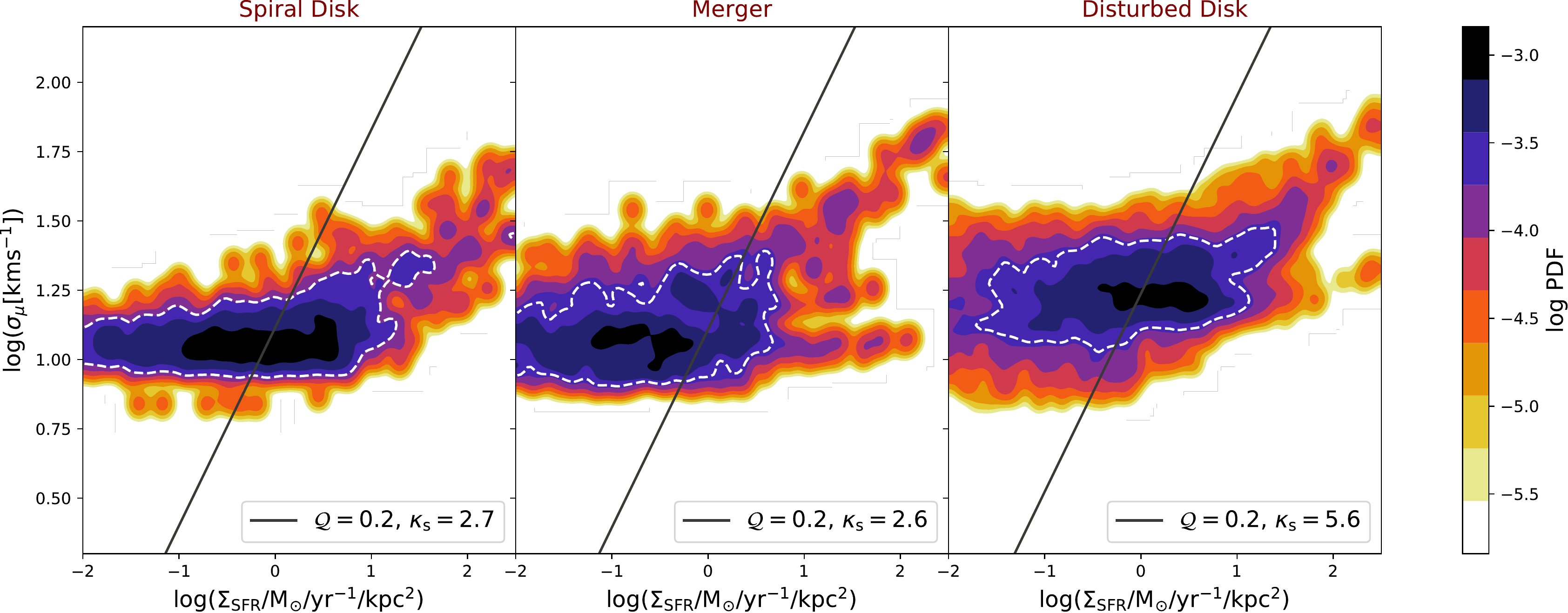}
\caption{2D PDFs of spatially resolved small-scale velocity dispersion and star formation rate densities for various stages of Freesia. The white dashed contours show the region including $90\%$ of the data. The solid lines indicate the analytical expression for $\sigma-\Sigma_{\rm{SFR}}$ relation (see eq. \ref{eq:sigma_SigmaSFR} and values given in Tab. \ref{tab:stages_properties}) 
\label{fig:sigma-mu-sfr-2d}
}
\end{figure*}
In this section, we analyze the small-scale velocity dispersion ($\sigma_{\mu}$) for three stages of Freesia. In Fig. \ref{fig:sigma-mu-pdf}, the PDFs of $\sigma_{\mu}/v_c$ in log-space are plotted for Freesia. We have fitted these profiles with a normal distribution:
\be
f(x)=\frac{A}{\sqrt{2\pi}\omega}\exp{(-(x-\beta)^2/2\omega^2)}\,.
\ee 
For all the three stages, we have been able to fit the overall distribution except for the tails. The tails are likely the result of recent starburst episodes for which turbulent energy has not yet had time to dissipate and reach a steady state. The fit parameters are reported in each plot. Among the stages, the PDF of the DD stage has a larger width which is due to the fact that the ISM is more turbulent in this stage and it has a broader thermal velocity distribution.
In Fig. \ref{fig:sigma-mu-sfr-2d}, the 2D PDFs of spatially resolved $\sigma_{\mu}$ as a function of $\Sigma_{\rm SFR}$ for three evolutionary stages are shown. We see that $\sigma_\mu$ in all the stages is almost constant independent of $\Sigma_{\rm SFR}$ over three orders of magnitude. The distribution of $\sigma_\mu$ is relatively narrow around the mean in all stages, with the DD one showing somewhat higher values. The $90\%$ data inclusion regions in these distributions lie on the derived average analytical expressions (see eq. \ref{eq:sigma_SigmaSFR}).

\bsp	
\label{lastpage}
\end{document}